\documentclass[preprint2]{aastex}
\usepackage{graphicx}
\usepackage{subfigure}

\shortauthors{Grillo et al.}

\begin{document}


\title{Golden gravitational lensing systems from the Sloan Lens ACS Survey. I. SDSS J1538+5817: one lens for two sources\altaffilmark{*,**}}

\altaffiltext{*}{Based on observations made with the NASA/ESA Hubble Space Telescope, obtained from the data archive at the Space Telescope Institute. STScI is operated by the association of Universities for Research in Astronomy, Inc. under the NASA contract NAS 5-26555.}
\altaffiltext{**}{Based on observations made with the Nordic Optical Telescope, operated
on the island of La Palma jointly by Denmark, Finland, Iceland,
Norway, and Sweden, in the Spanish Observatorio del Roque de los
Muchachos of the Instituto de Astrofisica de Canarias.}


\author{C. Grillo\altaffilmark{1,2}, T. Eichner\altaffilmark{2}, S. Seitz\altaffilmark{2,1}, R. Bender\altaffilmark{2,1}, M. Lombardi\altaffilmark{3,4}, R. Gobat\altaffilmark{5}, and A.~Bauer\altaffilmark{2}}
\email{cgrillo@usm.lmu.de}


\altaffiltext{1}{Max-Planck-Institut f\"ur extraterrestrische Physik, Giessenbachstr., D-85748 Garching bei M\"unchen, Germany}
\altaffiltext{2}{Universit\"ats-Sternwarte M\"unchen, Scheinerstr. 1, D-81679 M\"unchen, Germany}
\altaffiltext{3}{European Southern Observatory, Karl-Schwarzschild-Str. 2, D-85748 Garching bei M\"unchen, Germany}
\altaffiltext{4}{University of Milan, Department of Physics, via Celoria 16, I-20133 Milan, Italy}
\altaffiltext{5}{CEA Saclay, DSM/DAPNIA/Service d'Astrophysique, F-91191 Gif-sur-Yvette Cedex, France}


\begin{abstract}

We present a gravitational lensing and photometric study of the exceptional strong lensing system SDSS J1538+5817, identified by the Sloan Lens ACS survey. The lens is a luminous elliptical galaxy at redshift $z_{l}=0.143$. Using Hubble Space Telescope public images obtained with two different filters, the presence of two background sources lensed, respectively, into an Einstein ring and a double system is ascertained. Our new spectroscopic observations, performed at the Nordic Optical Telescope, reveal unequivocally that the two sources are located at the same redshift $z_{s}=0.531$. We investigate the total (luminous and dark) mass distribution of the lens between 1 and 4 kpc from the galaxy center by means of parametric and non-parametric lensing codes that describe the multiple images as point-like objects. Bootstrapping and Bayesian analyses are performed to determine the uncertainties on the quantities relevant to the lens mass characterization. Several disparate lensing models provide results that are consistent, given the errors, with those obtained from the best-fit model of the lens mass distribution in terms of a singular power law ellipsoid model. In particular, the lensing models agree on: (1) reproducing accurately the observed positions of the images; (2) predicting a nearly axisymmetric total mass distribution, centered and oriented as the light distribution; (3) measuring a value of $8.11^{+0.27}_{-0.59} \times 10^{10}\,M_{\odot}$ for the total mass projected within the Einstein radius of 2.5 kpc; (4) estimating a total mass density profile slightly steeper than an isothermal one [$\rho (r) \propto r^{-2.33^{+0.43}_{-0.20}}$]. A fit of the Sloan Digital Sky Survey multicolor photometry with composite stellar population models provides a value of $20^{+1}_{-4} \times 10^{10}\,M_{\odot}$ for the total mass of the galaxy in the form of stars and of $0.9^{+0.1}_{-0.2}$ for the fraction of projected luminous over total mass enclosed inside the Einstein radius. By combining lensing (total) and photometric (luminous) mass measurements, we differentiate the lens mass content in terms of luminous and dark matter components. This two-component modeling, which is viable only in extraordinary systems like SDSS J1538+5817, leads to a description of the global properties of the galaxy dark matter halo. Extending these results to a larger number of lens galaxies would improve considerably our understanding of galaxy formation and evolution processes in the $\Lambda$CDM scenario.

\end{abstract}





\section{Introduction}

Early-type galaxies host the majority of the baryonic mass observed in galaxies in the Universe (e.g., \citealt{fuk98}; \citealt{ren06}); hence, deciphering the processes that lead to their formation and the mechanisms that rule their subsequent evolution is a key cosmological issue. For instance, it is still debated whether early-type galaxies form at relatively high redshift ($z \lesssim 2$) as a result of a global starburst and then passively evolve to the present (e.g., \citealt{egg62}; \citealt{lar74}; \citealt{ari87}; \citealt{bre94}) or whether they assemble from mutual disruption of disks in merging events (e.g., \citealt{too77}; \citealt{whi78}). Information with which to distinguish these scenarios lies in the characteristics of galaxy dark-matter halos. However, the lack of suitable and easily interpreted kinematical tracers, such as HI in spirals, has made comprehensive studies on the dark matter component in early-type galaxies rather difficult (e.g., \citealt{ber92}; \citealt{sag92}; \citealt{tho07,tho09}).

In the last few years, strong gravitational lensing has allowed astrophysicists to make great progress in the understanding of the internal structure of early-type galaxies. Through lensing, it has become possible to address in detail some fundamental problems related to the mechanisms of formation of early-type galaxies, like the determination of the amount and distribution of dark matter (e.g., \citealt{gav07}; \citealt{gri08c,gri09}; \citealt{bar09}) or the investigation of the total mass density profile and its redshift evolution (e.g., \citealt{tre04}; \citealt{koo06}). Several algorithms have been developed in order to fit the observational data of a strong gravitational lens system and, thus, to reconstruct the properties of a lens.

Simplifying, a first difference between codes is the use of a parametric model (e.g., \emph{gravlens}\footnote{http://redfive.rutgers.edu/$\sim$keeton/gravlens/}, \citealt{kee01a,kee01b}; \citealt{sei98}; \citealt{war03}; \citealt{hal06}; \citealt{rze07}; \citealt{gri08c}) or a non-parametric model (e.g., \emph{PixeLens}\footnote{http://www.qgd.uzh.ch/projects/pixelens/}, \citealt{sah04}; \citealt{koo05}; \citealt{suy09}) to describe the mass distribution of a lens. In the former case, the mass distribution of a lens is assumed to be accurately described by an analytical expression; the fundamental scales of the model are determined by comparing the observed and model-predicted properties of the multiple images. In the latter case, a pixelated map or a multipole decomposition of the surface mass density of the lens is usually estimated through a statistical analysis that requires, in addition to the observational information, some extra physically plausible constraints, called priors, on the surface mass density distribution (e.g., positive-definite and smooth) of the lens. On the one hand, parametric models provide a great deal of freedom and complexity, but they do not cover ``naturally'' all the possible realistic mass distributions (for instance, surface mass density distributions with twisting isodensity contours); on the other hand, even if non-parametric models are more general, their number of degrees of freedom is often much larger than the constraints and this can result in three-dimensional density distributions that are dynamically unrealistic or unstable. A viable solution to obtain physically significant density distributions is to consider a framework where the mass distribution of the lens is reconstructed by combining in a fully self-consistent way both gravitational lensing and stellar dynamics measurements (e.g., \citealt{bar07}).

\begin{figure}[htb]
\centering
\includegraphics[width=0.4\textwidth]{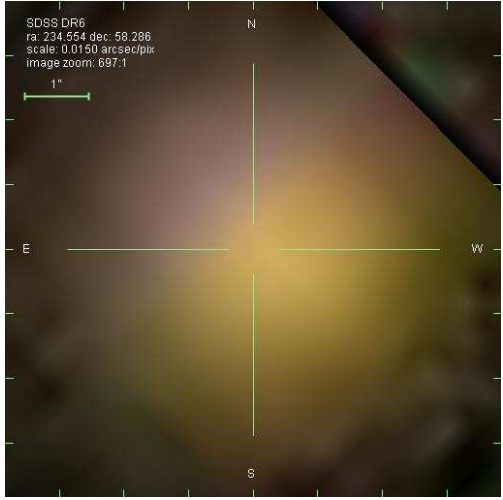}
\caption{SDSS \emph{gri} image centered on the lens galaxy.}
\label{fi01}
\end{figure}

\begin{figure}[htb]
\centering
\includegraphics[width=0.49\textwidth]{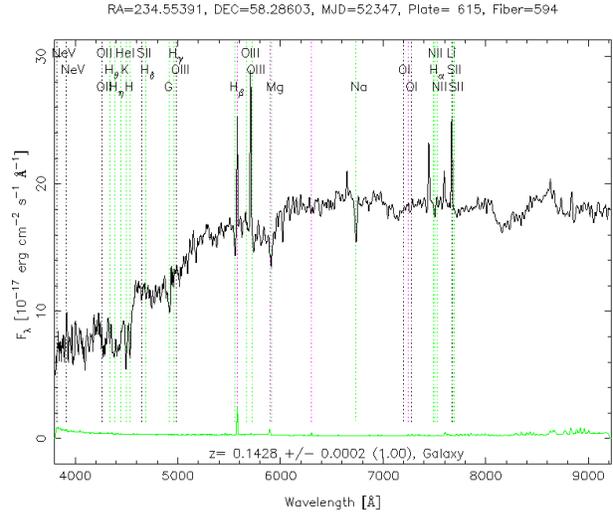}
\caption{SDSS spectrum obtained within an aperture of 3\arcsec $\,$ diameter centered on the lens galaxy.}
\label{fi03}
\end{figure}

Lensing codes are further distinguished by the fact that the multiple images and the corresponding sources can be modeled as point-like (e.g., \emph{gravlens}; \emph{PixeLens}; \citealt{sei98}; \citealt{hal06}) or extended (e.g., \citealt{war03}; \citealt{koo05}; \citealt{rze07}; \citealt{gri08c}; \citealt{suy09}) objects. In the context of point-like algorithms, the best-fit model is defined as that which minimizes the chi-square between the measured positions of the centroids of the images and the positions reproduced by the model, weighted by the measurement uncertainties. Additional chi-square terms that quantify the agreement between the observed and model-predicted relative fluxes and time delays of the multiple images can also be included. For extended algorithms, the goodness of a model is estimated by comparing on a pixelated grid the image surface brightness morphology and distribution which are observed to those which are reproduced by the model (after convolution with the relevant point spread function).

The relative positions of a multiply imaged system can sometimes be measured with an accuracy of a few milli-arcseconds (e.g., \citealt{pat01}) and these positions represent the most important constraints on the mass distribution of the lens. In fact, although the flux ratios of the multiple images can be easily estimated and offer another important source of information, the sensitivity of the flux measurements to details such as the dark matter substructure of the lens, the extinction in the interstellar medium of the lens, the microlensing effects of the stars present in the lens, and the time variability in the source decrease their potential. Time delays can also help to determine the mass distribution of a lens, but a statistically significant number of measurements of this kind is just starting to become available.

\begin{figure*}[htb]
\centering
\includegraphics[width=0.4\textwidth]{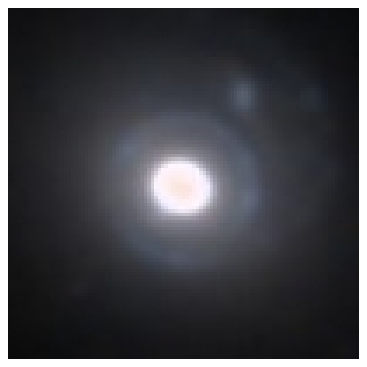}
\includegraphics[width=0.4\textwidth]{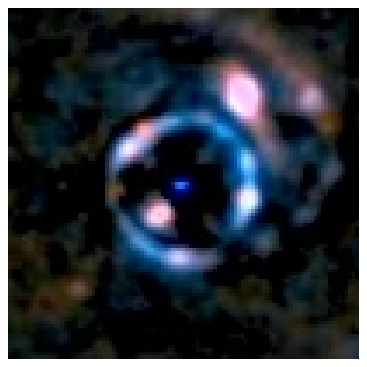}
\caption{Color images of a 5\arcsec $\times$ 5\arcsec $\,$ field around the gravitational lensing system SDSS J1538+5817, before (\emph{on the left}) and after (\emph{on the right}) the subtraction of an elliptical model, fitted on the luminosity profile of the lens galaxy. The images are obtained by combining the \emph{F606W} HST/WFPC2 and the \emph{F814W} HST/ACS filters.}
\label{fi02}
\end{figure*}

\begin{table*}
\centering
\caption{The lens galaxy.}
\begin{tiny}
\begin{tabular}{ccccccccccc}
\hline\hline \noalign{\smallskip}
RA & Dec & $z_{l}$ & $q_{\mathrm{L}}$ & $\theta_{q_{\mathrm{L}}}$ & $\theta_{e}$ & $u$ & $g$ & $r$ & $i$ & $z$ \\
(J2000) & (J2000) & & & (deg) & (\arcsec) & (mag) & (mag) & (mag) & (mag) & (mag) \\
\noalign{\smallskip} \hline \noalign{\smallskip}
15:38:12.92 & +58:17:09.8 & 0.143 & 0.82 & 157.3 & 1.58 & $19.50\pm0.06$ & $18.17\pm0.01$ & $17.17\pm0.01$ & $16.74\pm0.01$ & $16.43\pm0.01$ \\
\noalign{\smallskip} \hline
\end{tabular}
\end{tiny}
\begin{list}{}{}
\item[References --]\citealt{bol08}.
\item[Notes --]Magnitudes are extinction-corrected modelMag (AB) from the SDSS.
\end{list}
\label{tab1}
\end{table*}

The projected total mass enclosed within a cylinder of radius equal to the Einstein radius of a lensing system can be accurately measured by only fitting the observed positions of the multiple images (e.g., \citealt{koc91}; \citealt{gri08c}), whereas a detailed fit of the possible arcs and rings associated with an extended source is necessary if the interest is in the properties of both lens and source (e.g., \citealt{swi07}; \citealt{veg09}). By combining lensing and multiband photometric measurements, the amount of mass present in a lens galaxy in the forms of dark and visible matter can be determined (e.g., \citealt{gri08a,gri09}).

In addition to projected total mass, lensing analyses allow one to estimate also the total mass density profile of lens galaxies (e.g., \citealt{rus03}; \citealt{rus05}). This can be achieved either by combining in a statistical way lensing and stellar dynamics data in a sample of lens galaxies that are assumed to have a homologous structure (e.g., \citealt{koo06}), by performing a joint lensing and extended stellar kinematics study in a single lens galaxy (e.g., \citealt{bar09}; \citealt{tre04}), or by using lensing only in exceptional lensing systems that show multiple images of different sources probing wide angular and radial ranges of the lens mass distribution (e.g., \citealt{syk98}; \citealt{nai98}). 

In this paper, we study the lensing system SDSS J1538+5817, discovered by the Sloan Lens ACS (SLACS) survey\footnote{http://www.slacs.org/}. This system is particularly interesting because two different sources are lensed, one into an Einstein ring with four luminosity peaks and the other into two images, by an early-type galaxy that has an almost circular projected light distribution. The large number of images at various angular distances from the galaxy center and the nearly perfect axisymmetric lensing configuration of the ring makes this system the ideal laboratory to disentangle the luminous and dark components of the lens mass distribution.

The paper is organized as follows. In Sect.~2, we describe the observational data for the complex strong lensing system SDSS J1538+5817. We perform parametric lensing analyses of this system in Sect.~3. Then, in Sect.~4, we investigate the luminous and dark matter composition of the lens galaxy. In Sect.~5, we summarize the results obtained in this study. Finally, in the Appendix, we model the lens mass distribution on a pixelated grid and compare these non-parametric results to those from Sect.~3. Throughout this work we assume $H_{0}=70$ km s$^{-1}$ Mpc$^{-1}$, $\Omega_{\mathrm{m}}=0.3$, and $\Omega_{\Lambda}=0.7$. In this model, 1\arcsec $\,$ corresponds to a linear size of 2.51 kpc at the lens plane.

\section{Observations}

\begin{table}
\centering
\caption{Astrometric and photometric measurements for the multiple images.}
\begin{scriptsize}
\begin{tabular}{cccccccc}
\hline\hline \noalign{\smallskip}
 & $x^{\mathrm{a}}$ & $y^{\mathrm{a}}$ & $z_{s}$ & $\delta_{x,y}$ & $f$ & $\delta_{f}$ & $d^{\mathrm{a}}$ \\
 & (\arcsec) & (\arcsec) & & & (\arcsec) & & (\arcsec) \\
\noalign{\smallskip} \hline \noalign{\smallskip}
$D_{1}$ & 0.88 & 1.31 & 0.531 & 0.05 & 1.00 & 0.30 & 1.58 \\
$D_{2}$ & $-$0.33 & $-$0.40 & 0.531 & 0.05 & 0.17 & 0.09 & 0.52 \\
$Q_{1}$ & 0.96 & 0.33 & 0.531 & 0.05 & & & 1.02 \\
$Q_{2}$ & $-$0.75 & 0.60 & 0.531 & 0.05 & & & 0.96 \\
$Q_{3}$ & $-$0.15 & $-$0.98 & 0.531 & 0.05 & & & 0.99 \\
$Q_{4}$ & 0.95 & $-$0.19 & 0.531 & 0.05 & & & 0.97 \\
\noalign{\smallskip} \hline
\end{tabular}
\end{scriptsize}
\begin{list}{}{}
\item[$^{\mathrm{a}}$]With respect to the galaxy center.
\end{list}
\label{tab2}
\end{table}

\begin{figure}
\centering
\includegraphics[width=0.4\textwidth]{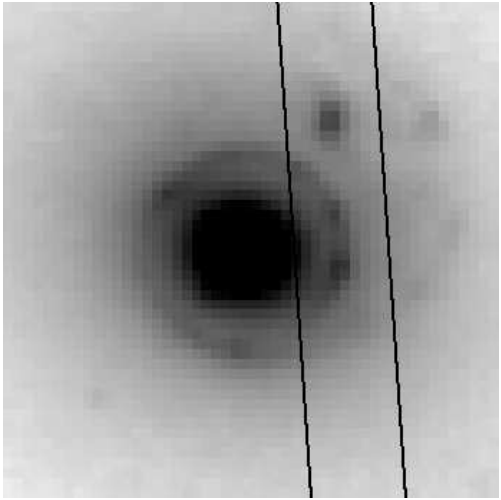}
\caption{Observational configuration for the spectroscopic measurements obtained at the NOT telescope with the ALFOSC instrument: position of the 1\arcsec $\,$long-slit.}
\label{fi22}
\end{figure}

\begin{figure*}
\centering
\includegraphics[width=\textwidth]{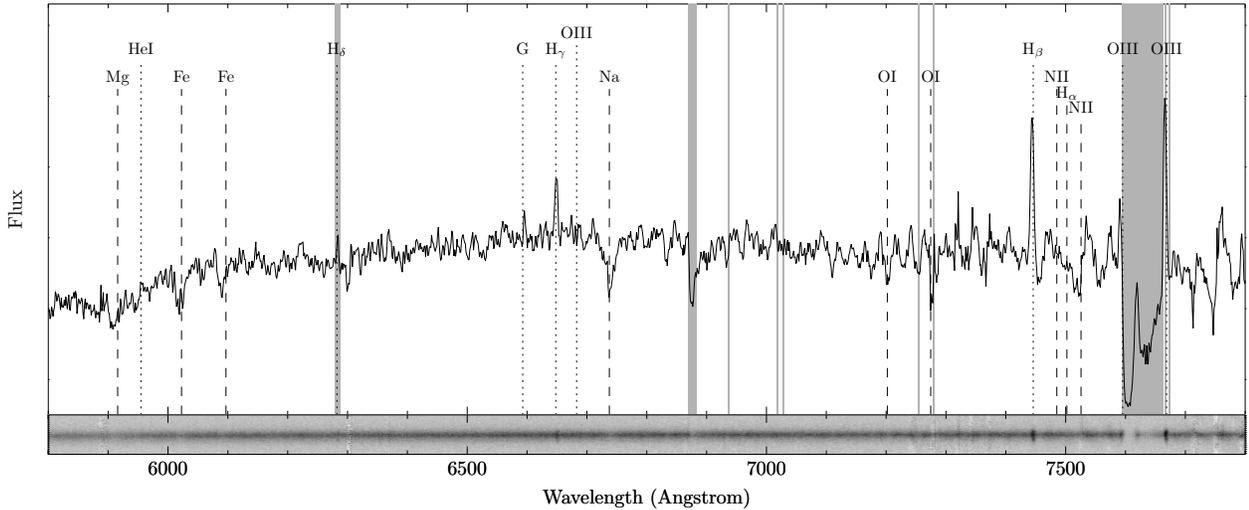}
\caption{The NOT/ALFOSC 1D (\emph{on the top}) and 2D (\emph{on the bottom}) spectra of the lensing system SDSS J1538+5817. In the wavelength range shown here, the most prominent spectral features at the redshift of the lens ($z_{l}=0.143$) and sources ($z_{s}=0.531$) are marked, respectively, with dashed and dotted lines. The shaded regions indicate the principal telluric lines. The flux is given in arbitrary units.}
\label{fi20}
\end{figure*}

\begin{figure}
\centering
\includegraphics[width=0.49\textwidth]{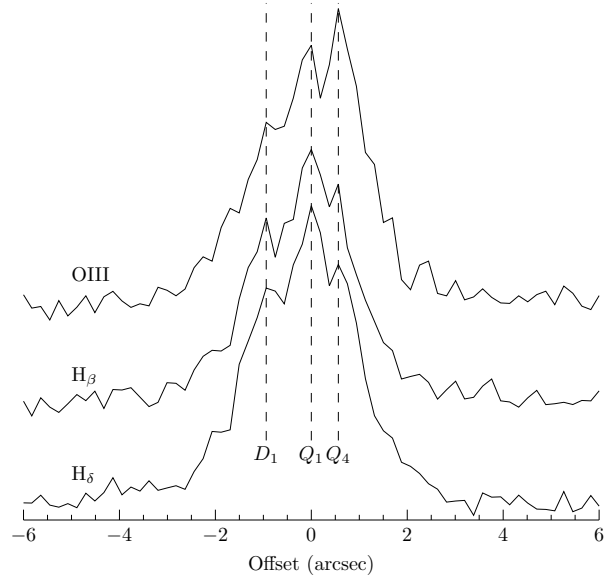}
\caption{Intensity (in arbitrary units) of the three $H_{\delta}$, $H_{\beta}$, and [OIII] $\lambda 5007$ emission lines along the slit width. The dashed lines mark the positions of three observed peaks that trace closely the geometrical configuration of the $D_{1}$, $Q_{1}$, and $Q_{4}$ images.}
\label{fi21}
\end{figure}

The SLACS survey was started in 2003 and aims at studying, from a lensing and dynamics perspective, a statistically significant number of galaxies acting as strong lenses and located at redshifts lower than 0.5. The candidate lenses were spectroscopically selected from the Sloan Digital Sky Survey (SDSS)\footnote{http://www.sdss.org/} database by identifying those objects that show, in addition to the continuum and absorption lines of a possible lens galaxy at redshift $z_{l}$, one or multiple emission lines of a hypothetical source at a higher redshift $z_{s}$. The most promising candidates were then observed at least once with the \emph{Hubble Space Telescope} (HST) Advanced Camera for Survey (ACS) to confirm the lens hypothesis (for further information, see \citealt{bol06,bol08}). This procedure resulted in the sample of 63 ``grade-A'' strong gravitational lensing systems presented in \citet{bol08}. SDSS J$1538+5817$ is one of the lens galaxies discovered by the SLACS survey. The photometric and spectroscopic observations taken by the SDSS are shown in Figs. \ref{fi01} and \ref{fi03}. As described above, the redshifts of the lens galaxy and a source ($z_{l}=0.143$ and $z_{s}=0.531$) were measured from different spectral features. 

By making use of the publicly available observations in the \emph{F814W} and \emph{F606W} filters of the HST/ACS and Wide Field Planetary Camera 2 (WFPC2) respectively, we model the luminosity distribution of the lens galaxy and subtract the best-fit model from the images. In detail, a model for the lens galaxy is constructed by using an iterative procedure: first, presumable background source images are masked and isophotal contours of the lens galaxy are derived for surface brightness levels separated by 0.1 mag arcsec$^{-2}$. Then, all isophote contours (even if partially masked) are fitted by ellipses following the method of \citet{ben87}. This provides five parameters (center coordinates, major and minor axis, and position angle) per surface brightness level. The resulting table of these parameters and associated surface brightnesses is employed to calculate a smooth elliptical model for the lens galaxy. This model is subtracted from the image, leaving as residuals only the images of the background source galaxy. The residuals are used to check and improve the masking of the source galaxy. This procedure is repeated twice until the final model for the lens galaxy is obtained. In addition, images taken in different wavelength bands provide color information for lens and source which is of additional help to identify and separate lens and source components. Color composite images of the strong lensing system and the residuals after the lens galaxy subtraction are shown in Fig. \ref{fi02}.

The excellent angular resolution of the HST allows us to identify two systems of multiple images (for labels, see Fig. \ref{fi04}): a double ($D_{1}$ and $D_{2}$) and a quad ($Q_{1}$, $Q_{2}$, $Q_{3}$, and $Q_{4}$). The images of the two systems have different colors, but the average distances from the galaxy center of $D_{1}$ and $D_{2}$ and $Q_{1}$, $Q_{2}$, $Q_{3}$, and $Q_{4}$ are consistent within the errors. This fact would imply approximately the same redshift for the two sources, if the lens total mass were close to an isothermal distribution. In addition, the absence in the SDSS spectrum of evident emission lines at a possible third redshift supports the hypothesis that the two sources are at the same distance to the observer.

Since the precise knowledge of the redshift of the two sources plays a crucial role in the determination of the total mass distribution of the lens galaxy, we decided to perform additional spectroscopic measurements to understand whether the emission lines observed in the SDSS spectrum are associated to one or both of the lensed sources. The data were obtained on June 25, 2009 as a Fast-Track Observing Program (P38-428) with the Andalucia Faint Object Spectrograph and Camera (ALFOSC) at the 2.5-m Nordic Optical Telescope (NOT) on La Palma (Spain). We positioned a 1\arcsec-wide long-slit centered in $Q_{1}$ and passing through $D_{1}$ and $Q_{4}$, as shown in Fig. \ref{fi22}. We used ALFOS\-C with the 8 grism, that covers a wavelength range between 5825 and 8350 \AA $\,$with a dispersion of 1.3 \AA $\,$per pixel. In good atmospheric conditions (seeing between 0.7 and 1\arcsec) and in the same observational configuration, we obtained six exposures of 24 minutes each, resulting in a total integration time of 2.4 hrs.

In Fig. \ref{fi20}, we show the wavelength-calibrated 1D and 2D spectra. We identify several prominent absorption lines at redshift 0.143 and at least six secure emission lines (H$_{\delta}$, G, H$_{\gamma}$, H$_{\beta}$, [OIII] $\lambda 4959$, and [OIII] $\lambda 5007$) at redshift 0.531. At the emission line positions, the presence of two intensity peaks, below and above the continuum, is visible in the 2D spectrum. In Fig. \ref{fi21}, we plot the intensity of the H$_{\delta}$, H$_{\beta}$, and [OIII] $\lambda 5007$ emission lines as a function of spatial position along the slit's cross-section. This corresponds to a representation of three sections of the 2D spectrum at the emission line abscissas. For these three emission lines, we distinguish three intensity peaks located, with respect to the continuum (lying between $Q_{1}$ and $Q_{4}$), at angular positions consistent with those of the three images $D_{1}$, $Q_{1}$, and $Q_{4}$. The measurement of the same emission lines at the same observed wavelengths proves in a conclusive way that the two sources $D$ and $Q$ are equally distant from the observer. We remark that the intensity values of the emission lines shown in Fig. \ref{fi21} are differently contaminated by the lens galaxy flux.

In Table \ref{tab1}, we summarize the photometric and spectroscopic properties of the lens galaxy: the coordinates (RA, Dec, $z_{l}$), the minor to major axis ratio ($q_{\mathrm{L}}$) and its position angle ($\theta_{q_{\mathrm{L}}}$, degrees east of north), and the SDSS multiband magnitudes ($u$, $g$, $r$, $i$, $z$). In Table \ref{tab2}, we report the coordinates of the multiple images ($x$, $y$, $z_{s}$) and the adopted position uncertainty on the first two coordinates ($\delta_{x,y}$), the relative flux of the double system components ($f$) and the respective error ($\delta_{f}$), and the distance of the images to the galaxy center ($d$).

\section{Strong gravitational lensing}

We address parametric (Sect. \ref{sec:parametric models}) point-like modeling of the strong gravitational lensing system. We focus mainly on projected total mass and total mass density profile measurements. A comparison with the results obtained from non-parametric models is provided in the Appendix.

\subsection{Parametric models}
\label{sec:parametric models}

\begin{table*}
\centering
\caption{The best-fit (minimum $\chi^{2}$) parameters of the different models.}
\begin{tabular}{ccccccccccc}
\hline\hline \noalign{\smallskip}
Model & $b$ & $x_{l}$ & $y_{l}$ & $q$ & $\theta_{q}$ & $\theta_{e}$ & $\gamma$ & $\chi^2$ & d.o.f. \\
 & (\arcsec) & (\arcsec) & (\arcsec) & & (deg) & (\arcsec) & & & \\
\hline
deV (nf) & 1.97 & $-$0.02 & 0.03 & 0.782 & 148.0 & 1.58 & & 1.02 & 3 \\
deV (wf) & 1.83 & $-$0.02 & 0.02 & 0.871 & 147.9 & 1.58 & & 2.93 & 4 \\
SIE (nf) & 1.08 & $-$0.03 & 0.04 & 0.866 & 147.5 & & 2.00 & 1.15 & 3 \\
SIE (wf) & 1.05 & $-$0.02 & 0.02 & 0.919 & 147.6 & & 2.00 & 3.77 & 4 \\
PL (nf) & 0.82 & $-$0.02 & 0.03 & 0.800 & 147.9 & & 2.33 & 0.36 & 2 \\
PL (wf) & 0.71 & $-$0.01 & 0.02 & 0.820 & 148.0 & & 2.47 & 0.99 & 3 \\
\noalign{\smallskip} \hline
\end{tabular}
\begin{list}{}{}
\item[Notes --]The notation (wf) and (nf) indicates, respectively, if the flux measurements of the double system are included in the modeling or not.
\end{list}
\label{tab3}
\end{table*}

\emph{Gravlens} (\citealt{kee01a}) is a publicly-available lensing software that, starting from the measured observables of a strong lensing system, reconstructs the properties of a lens in terms of an adopted model that is defined by some relevant parameters. By using this code, we perform a parametric analysis in which we describe the total mass distribution of the lens galaxy in terms of either an elliptical de Vaucouleurs model (deV), or a singular isothermal ellipsoid (SIE) model, or a singular power law ellipsoid (PL) model (for further details on the model definitions, see e.g. \citealt{kee01b}). Both a deV and an SIE model are characterized by five parameters: a length scale $\tilde{b}$ (corresponding to the value of the Einstein angle $\theta_{\mathrm{Ein}}$ in the circular limit), the two coordinates of the center $(x_{l},y_{l})$, the minor to major axis ratio $q$, and its position angle $\theta_{q}$. For the deV model, we fix the value of the effective angle ($\theta_{e}$) to that shown in Table \ref{tab1}. A PL model is more general than an SIE model. In particular, the former requires as an additional parameter the value of the exponent $\gamma$ of the three dimensional density distribution $\rho(r) \propto r^{-\gamma}$ (an SIE model is retrieved by setting $\gamma$ equal to 2). The convergence $\kappa(x,y)$ of a PL model, defined as the surface mass density of the model divided by the critical surface mass density of the studied lensing system (for definitions, see \citealt{sch92}), depends on the previous parameters as follows
\begin{equation}
\kappa (x,y) \propto \frac{\tilde{b}^{\gamma-1}}{\Big( x^2+\frac{y^2}{q^2} \Big) ^{\frac{\gamma-1}{2}}}\,.
\label{eq01}
\end{equation}
Due to the normalization used in the code, \emph{gravlens} provides values of a length scale $b$ that are related to the values of $\tilde{b}$ by a function $f(\cdot)$ of the axis ratio $q$:
\begin{equation}
b = \tilde{b}\,f(q)\,.
\label{eq02}
\end{equation}
Varying the parameters of the two adopted mass models and the positions of the sources [$(x_{D},y_{D});(x_{Q},y_{Q})$], we minimize a chi-square $\chi^{2}$ function. This function compares first only the observed (see Table \ref{tab2}) and predicted positions of the multiple images [deV(nf), SIE (nf), and PL (nf) models] and then also the measured (see Table \ref{tab2}) and reconstructed fluxes of the double system [deV (wf), SIE (wf), and PL (wf) models]. In the latter case, the flux of the source imaged twice is an additional free parameter of the models. In our lensing analysis, we decide to neglect the flux constraints relative the quad system because the presence of the Einstein ring prevents us from separating accurately the individual components. For the multiple images, we assume position uncertainties equal to the size of one pixel of ACS (0.05$\arcsec$) and flux uncertainties as reported in Table \ref{tab2} and determined by considering the different level of contamination on the flux estimates by the surface brightness of the lens galaxy. 

To estimate the statistical errors in the parameters characterizing each model, we perform 2000 $\chi^{2}$ minimizations on simulated data sets. These are obtained by extracting the image positions and fluxes from Gaussian distributions centered on the measured values and with standard deviations equal to the observational errors reported in Table \ref{tab2}. In addition, starting from the sets of optimized parameter values, we estimate the total projected mass $M_{\mathrm{len}}^{\mathrm{tot}}(\leq R_{i})$ enclosed within seven different circular apertures of radii $R_{i}$. The first three radii are chosen as the projected distances from the lens galaxy center of the inner image of the double system, the ``average'' Einstein circle, and the outer image of the double system. The remaining four radii are given by the values of the midpoints of the three segments defined by the previous three points and a further point at the same distance from the outer double image as the first point is from the second one [i.e., (0.26, 0.52, 0.75, 0.98, 1.28, 1.58, 1.84)\arcsec]. 

The best-fit (minimum chi-square) parameter and $\chi^{2}$ values of the different models are summarized in Table \ref{tab3}. For all the models, we find that the best-fit $\chi^{2}$ values are smaller than the corresponding number of degrees of freedom (d.o.f.). This implies that the reconstructed positions of the images are angularly very close to the measured positions of Table \ref{tab2}. In this section, we concentrate on the results given by the one-component PL and, as a comparison with the results of previous studies, simpler SIE models and only in the next section we will address the two-component mass decomposition.

\begin{figure}[htb]
\centering
\includegraphics[width=0.22\textwidth]{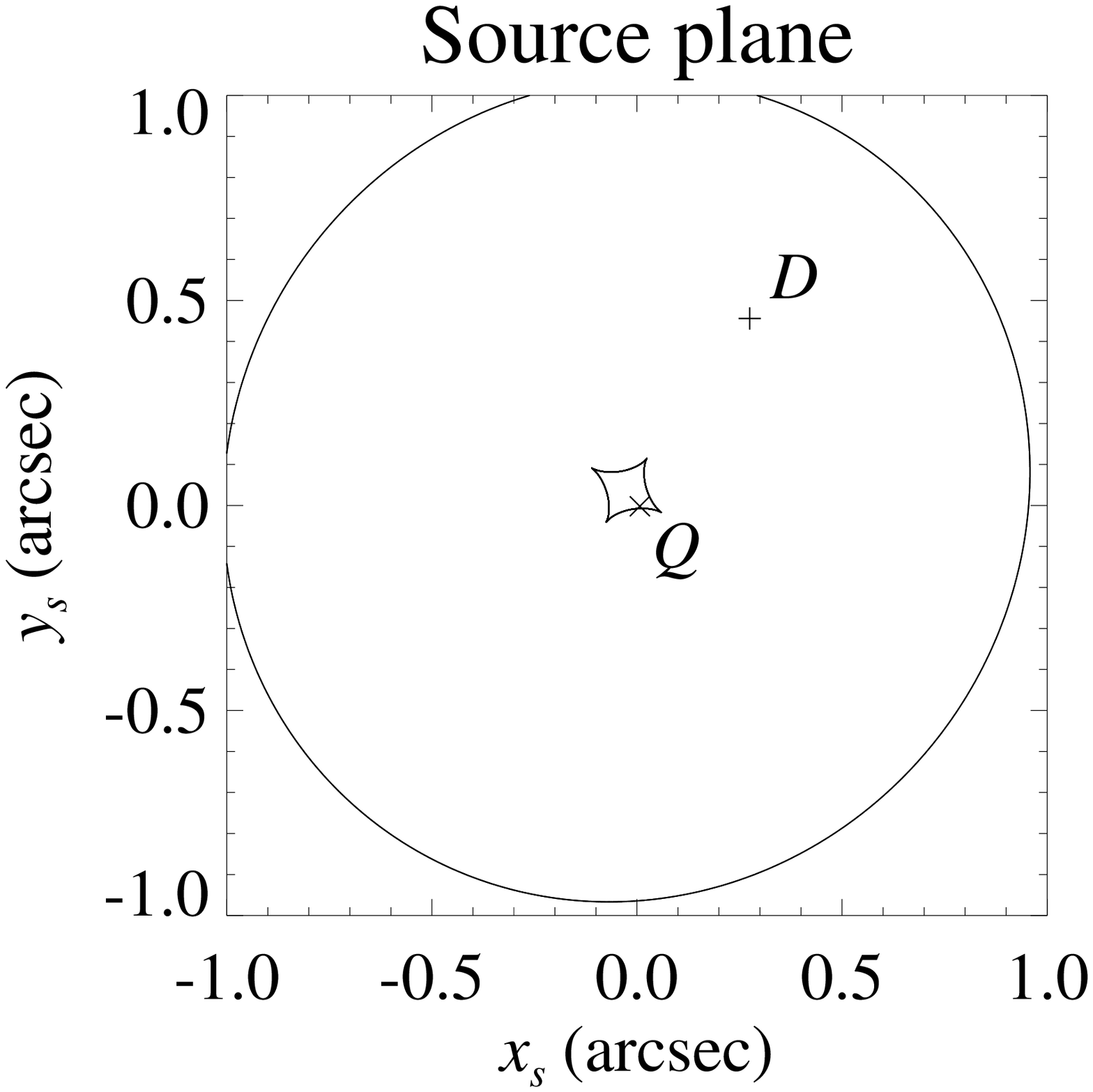}
\includegraphics[width=0.22\textwidth]{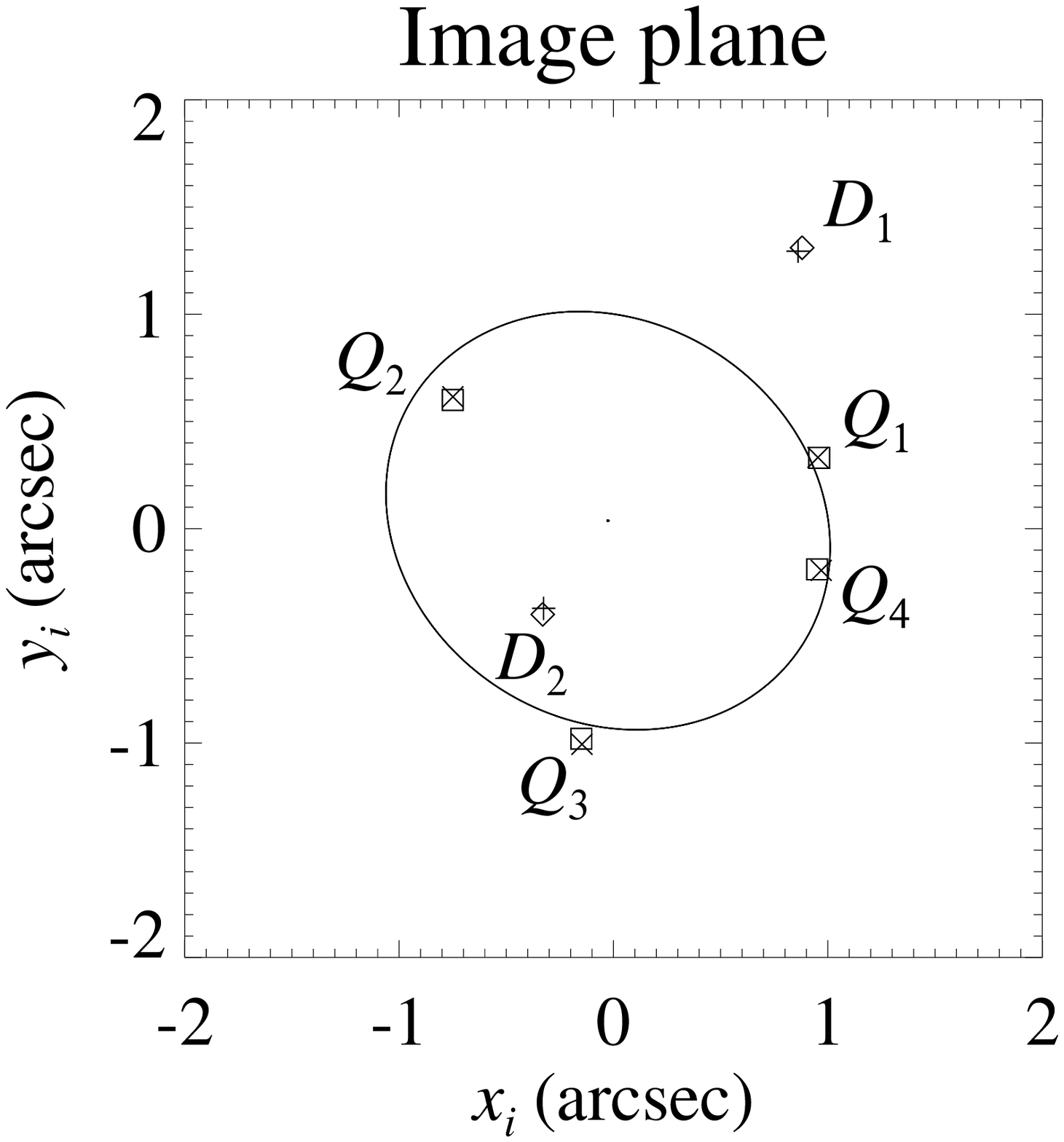}
\includegraphics[width=0.22\textwidth]{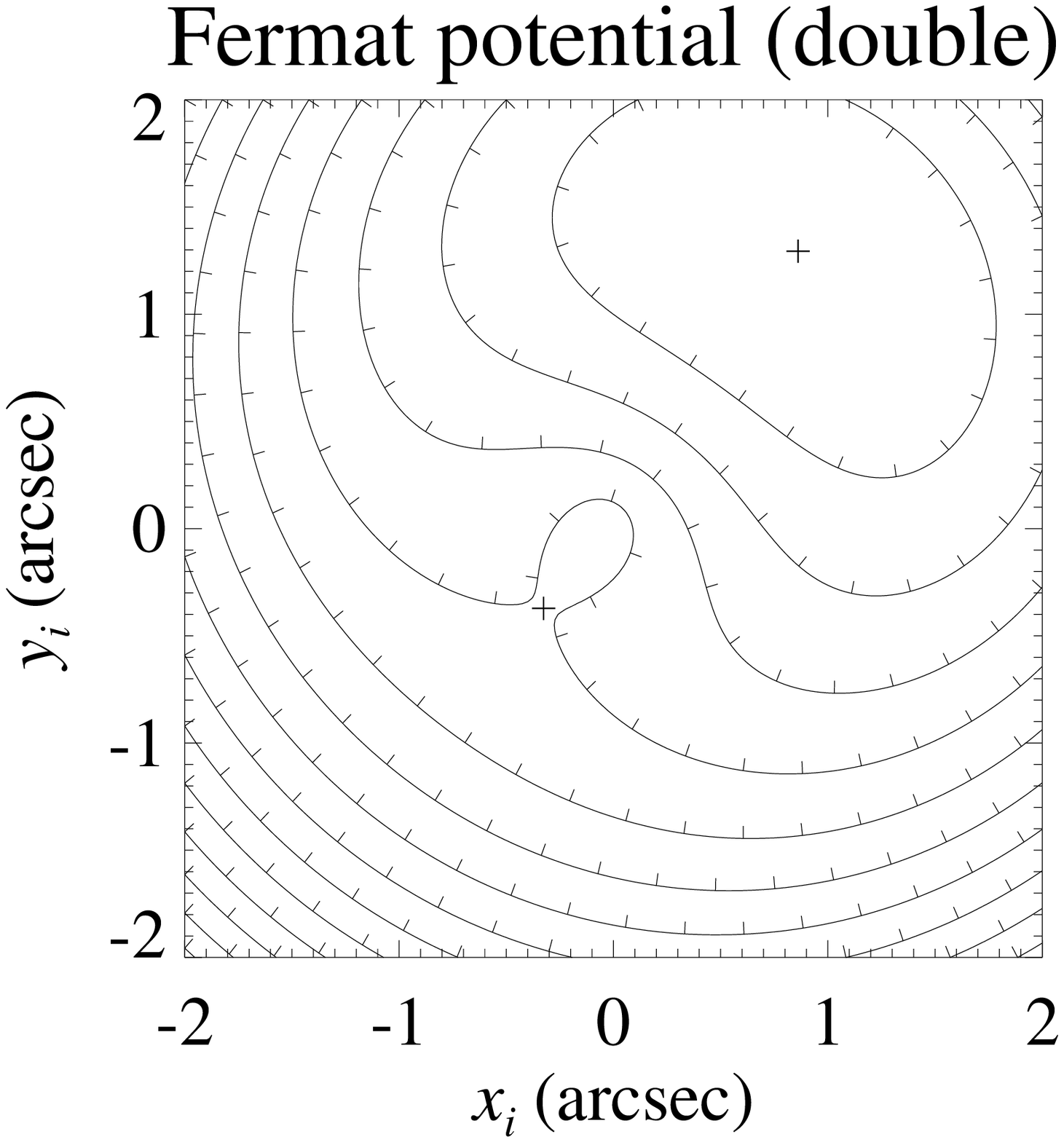}
\includegraphics[width=0.22\textwidth]{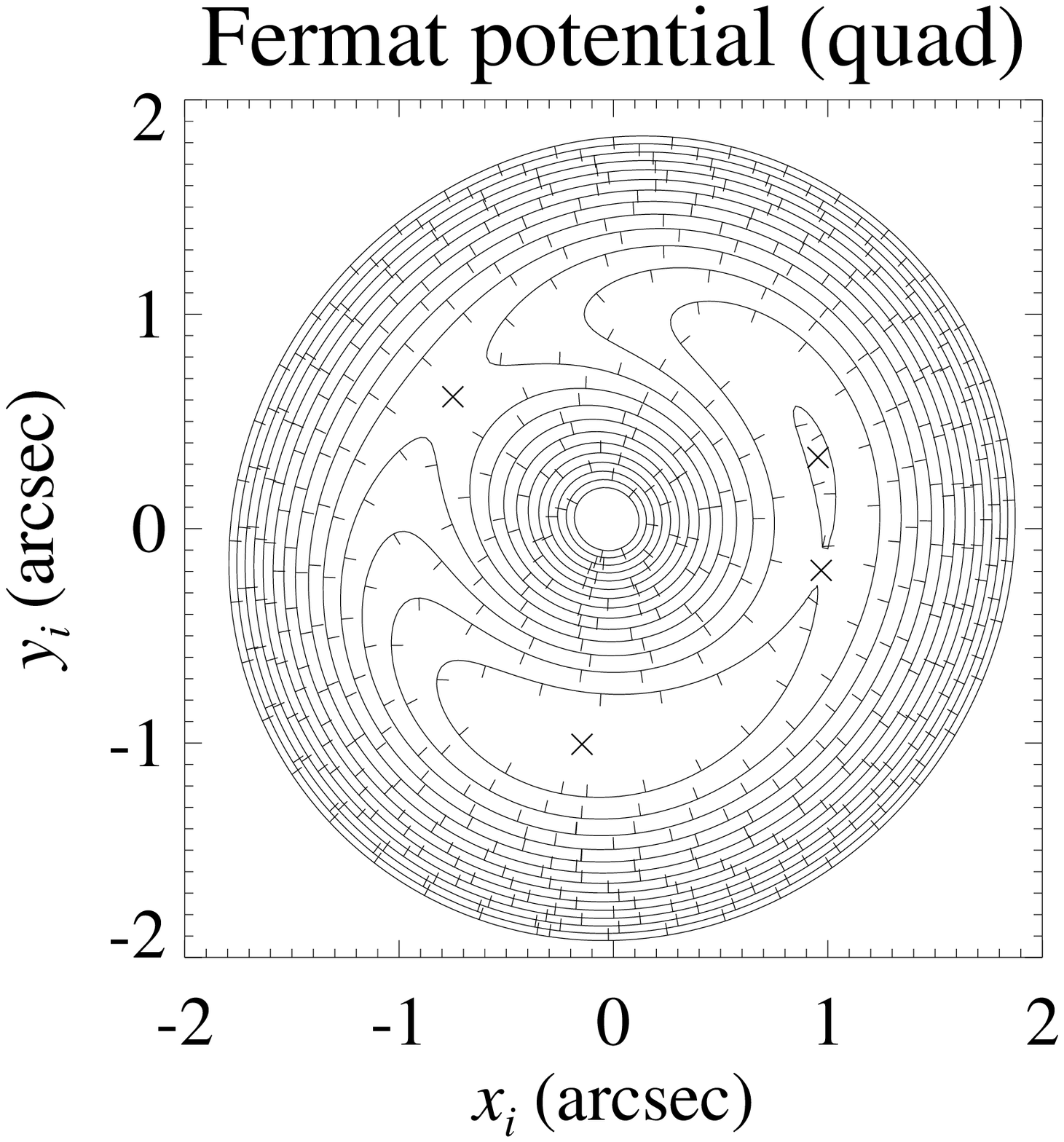}
\caption{Best-fit SIE (nf) model. \emph{Top left}: source plane with caustics. The predicted source positions of the double and quad systems are represented by a plus and a cross symbol, respectively. \emph{Top right}: image plane with critical curves. The observed and predicted image positions of the double (diamond and plus symbols, respectively) and quad (square and cross symbols, respectively) systems are shown. \emph{Bottom left}: contour levels of the Fermat potential for the double system. The images are one minimum (D$_{1}$) and one saddle point (D$_{2}$). \emph{Bottom right}: contour levels of the Fermat potential for the quad system. The images are two minima (Q$_{1}$ and Q$_{3}$) and two saddle points (Q$_{2}$ and Q$_{4}$).}
\label{fi04}
\end{figure}

In Fig. \ref{fi04}, for the best-fit SIE (nf) model we show the reconstructed positions of the sources and the caustics, the observed and reconstructed positions of the images and the critical curves, and the Fermat potential (for definition, see \citealt{sch92}) with its stationary points. The inclusion of the fluxes of the double system does not change significantly the best-fit parameters of the models. Moreover, the projected total mass estimates, which are presented below, are not particularly sensitive to the flux constraints. For these reasons, in the following we will mainly concentrate on the properties of the models that omit the additional source of information coming from the fluxes of the double system, i.e., the SIE (nf) and PL (nf) models. We note that the best-fit $b$ values are on the order of 1\arcsec, the typical distance of an image of the quad system from the center of the lens (see Tables \ref{tab2} and \ref{tab3}). The best-fit values of the lens center and ellipticity show that the lens mass distribution is well centered and aligned with the galaxy light distribution (see Tables \ref{tab1} and \ref{tab3}). In particular, we remark that the total surface mass of the lens is well approximated by an axisymmetric distribution. The best-fit values of the parameter $\gamma$ of the PL models suggest that the lens total density profile is slightly steeper than an isothermal one. We estimate maximum time delays of approximately 30 and 3 days for the double and quad systems, respectively (see Table \ref{tab4} and Fig. \ref{fi04}). Finally, as far as the positions of the sources are concerned, the model predicted angular distance of the two sources is between approximately 0.5 and 0.7\arcsec (see Table \ref{tab5}), corresponding, respectively, to 3.2 and 4.5 kpc at a redshift of 0.531. 

\begin{table}
\centering
\caption{The model-predicted time delays for the best-fit model parameters given in Table \ref{tab3}.}
\begin{tabular}{ccccc}
\hline\hline \noalign{\smallskip}
Model & $\Delta t_{D_{2,1}}$ & $\Delta t_{Q_{1,3}}$ & $\Delta t_{Q_{4,3}}$ & $\Delta t_{Q_{2,3}}$ \\
 & (days) & (days) & (days) & (days) \\ 
\noalign{\smallskip} \hline
SIE (nf) & 24.8 & 0.66 & 0.70 & 2.88 \\
PL (nf) & 33.5 & 0.82 & 0.87 & 3.26 \\
\noalign{\smallskip} \hline
\end{tabular}
\label{tab4}
\end{table}

\begin{table}
\centering
\caption{The model-predicted source positions for the best-fit model parameters given in Table \ref{tab3}.}
\begin{tabular}{cccc}
\hline\hline \noalign{\smallskip}
Model & $(x_{D},y_{D})$ & $(x_{Q},y_{Q})$ & $d_{D,Q}$ \\
 & (\arcsec,\arcsec) & (\arcsec,\arcsec) & (\arcsec) \\ 
\noalign{\smallskip} \hline
SIE (nf) & (0.27,0.46) & (0.00,0.00) & 0.53 \\
PL (nf) & (0.39,0.60) & (0.02,$-0.01$) & 0.72 \\
\noalign{\smallskip} \hline
\end{tabular}
\label{tab5}
\end{table}

\begin{figure*}[!htb]
\centering
\includegraphics[width=\textwidth]{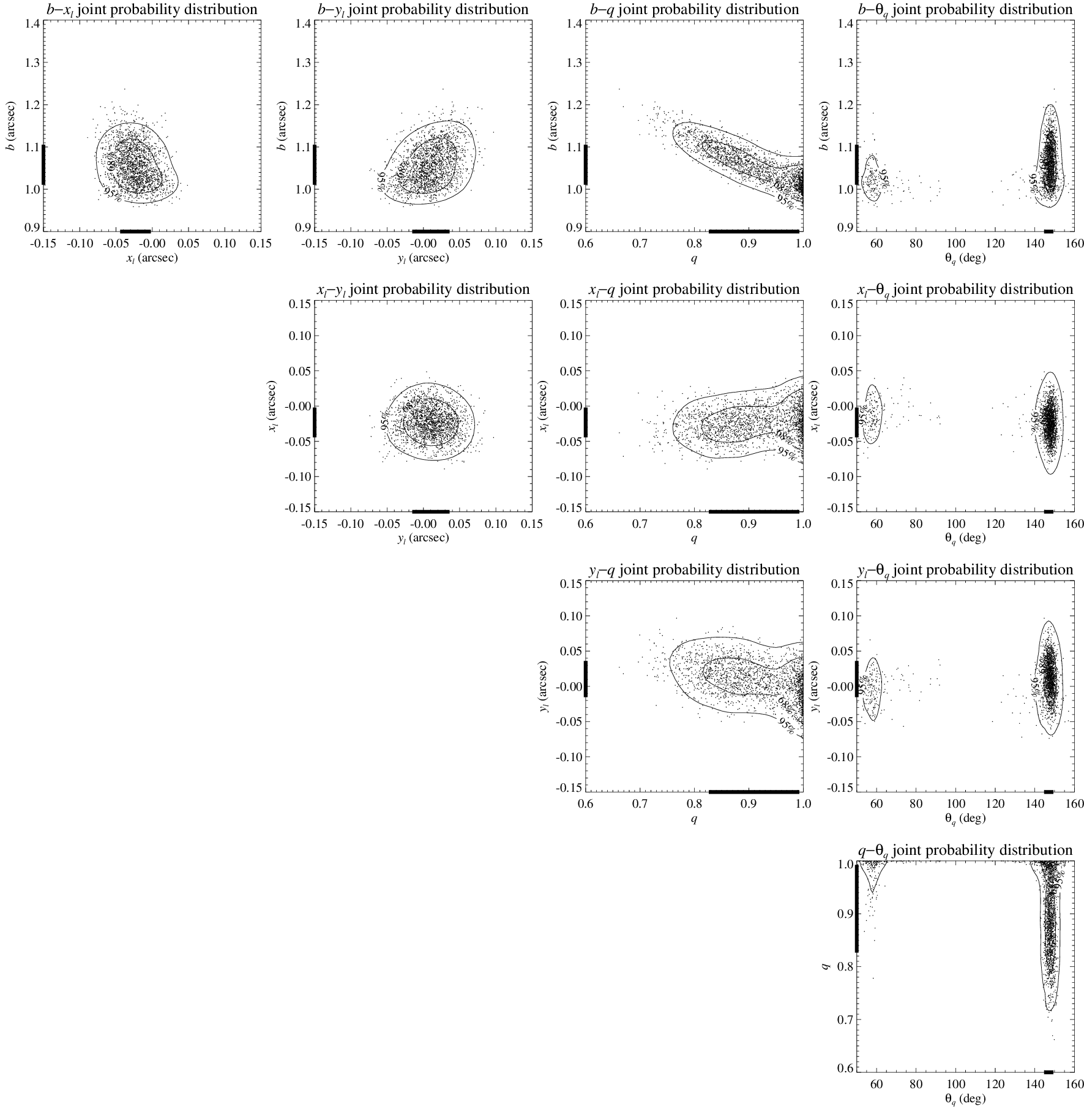}
\caption{Estimates of the errors and correlations in the parameters for an SIE (nf) model. Results of the $\chi^{2}$ minimizations on 2000 Monte-Carlo simulated data sets. Thick bars on the co-ordinate axes and contour levels on the planes represent, respectively, the 68\% confidence intervals and the 68\% and 95\% confidence regions. For each model parameter, the 68\% confidence interval is determined by excluding from the 2000 $\chi^{2}$ minimizations the 320 smallest and the 320 largest values.}
\label{fi05}
\end{figure*}

\begin{figure*}[!htb]
\centering
\includegraphics[width=\textwidth]{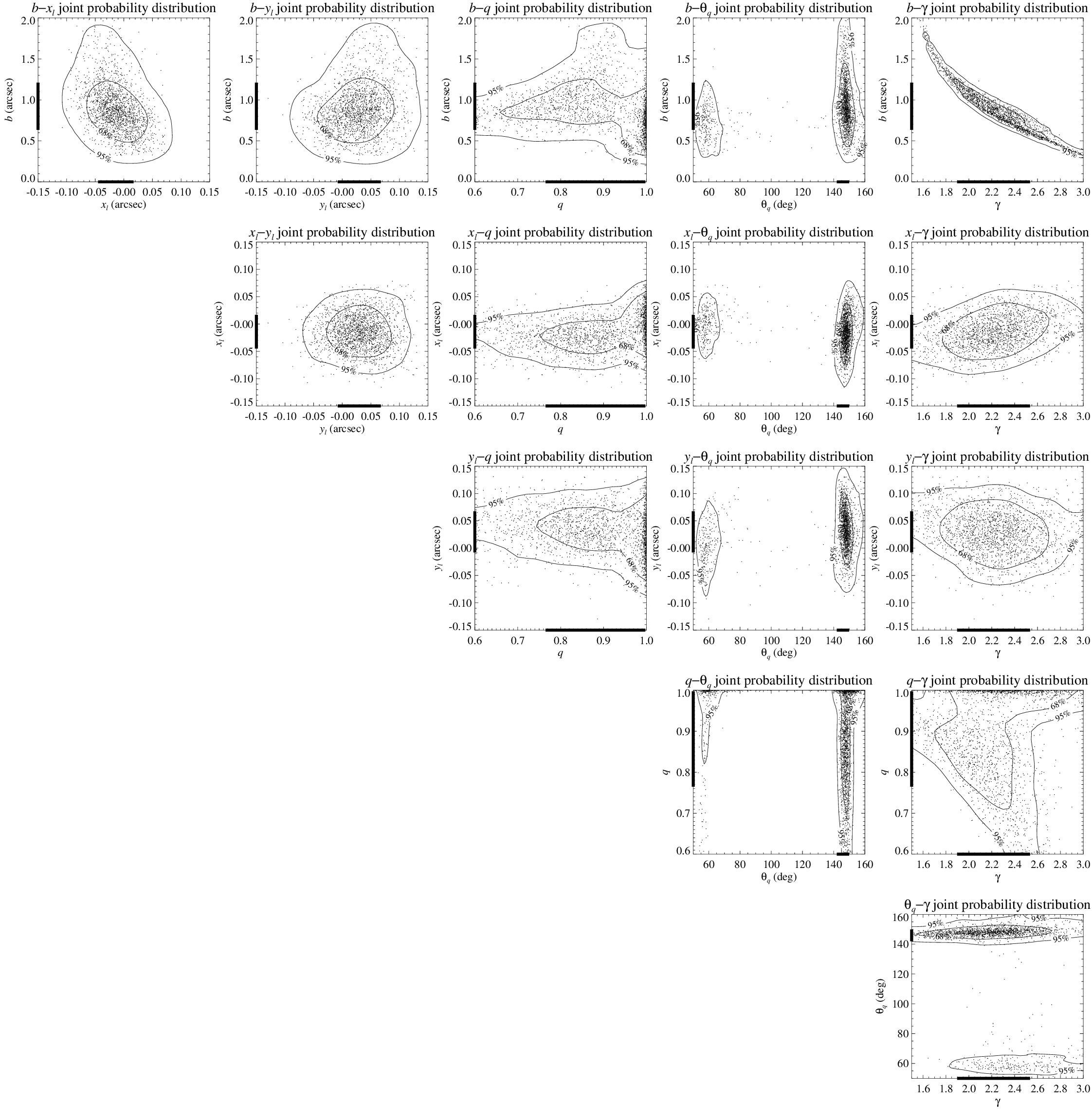}
\caption{Estimates of the errors and correlations in the parameters for a PL (nf) model. Results of the $\chi^{2}$ minimizations on 2000 Monte-Carlo simulated data sets. Thick bars on the co-ordinate axes and contour levels on the planes represent, respectively, the 68\% confidence intervals and the 68\% and 95\% confidence regions. For each model parameter, the 68\% confidence interval is determined by excluding from the 2000 $\chi^{2}$ minimizations the 320 smallest and the 320 largest values.}
\label{fi06}
\end{figure*}

\begin{figure}[htb]
\centering
\includegraphics[width=0.22\textwidth]{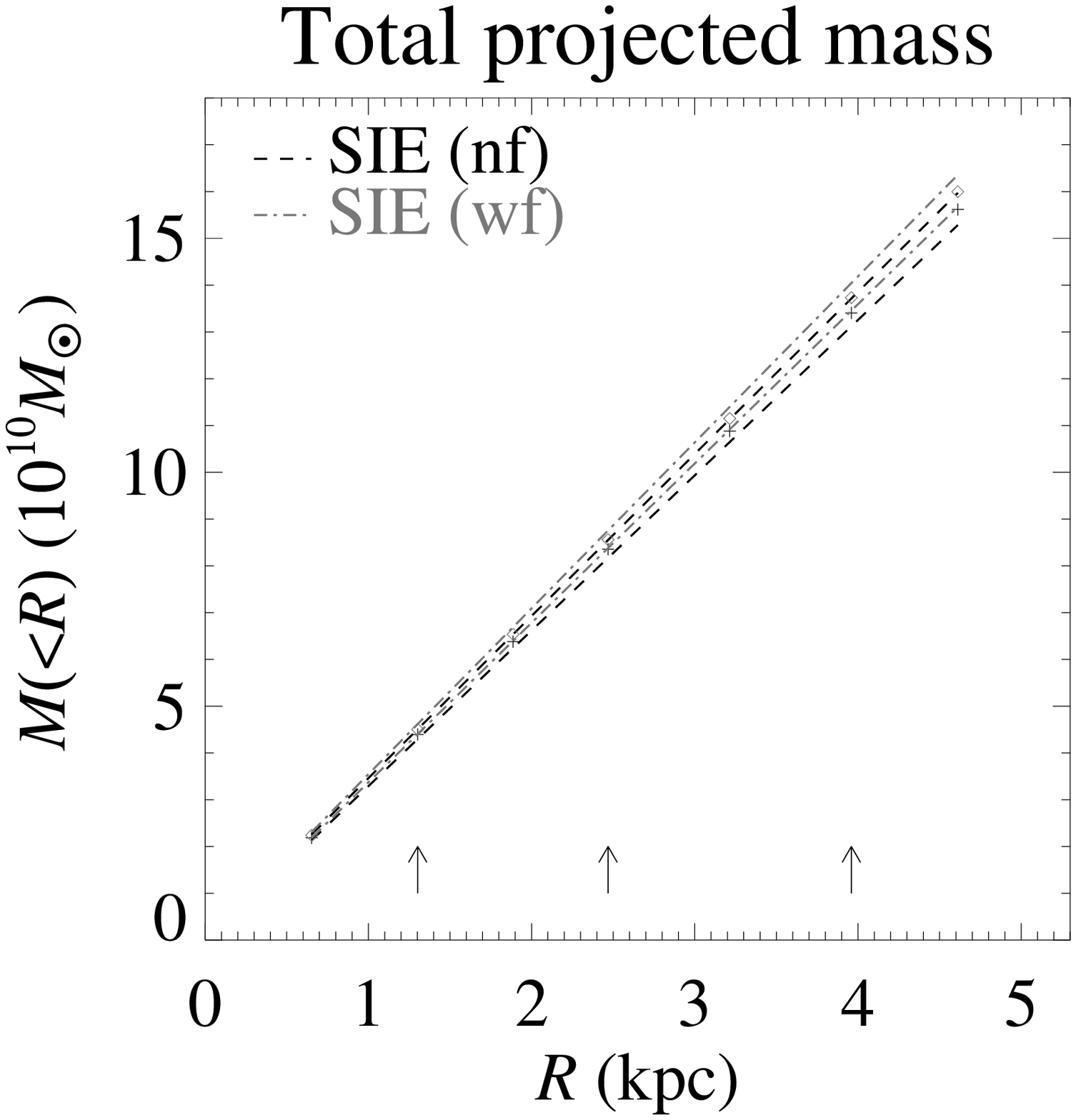}
\includegraphics[width=0.22\textwidth]{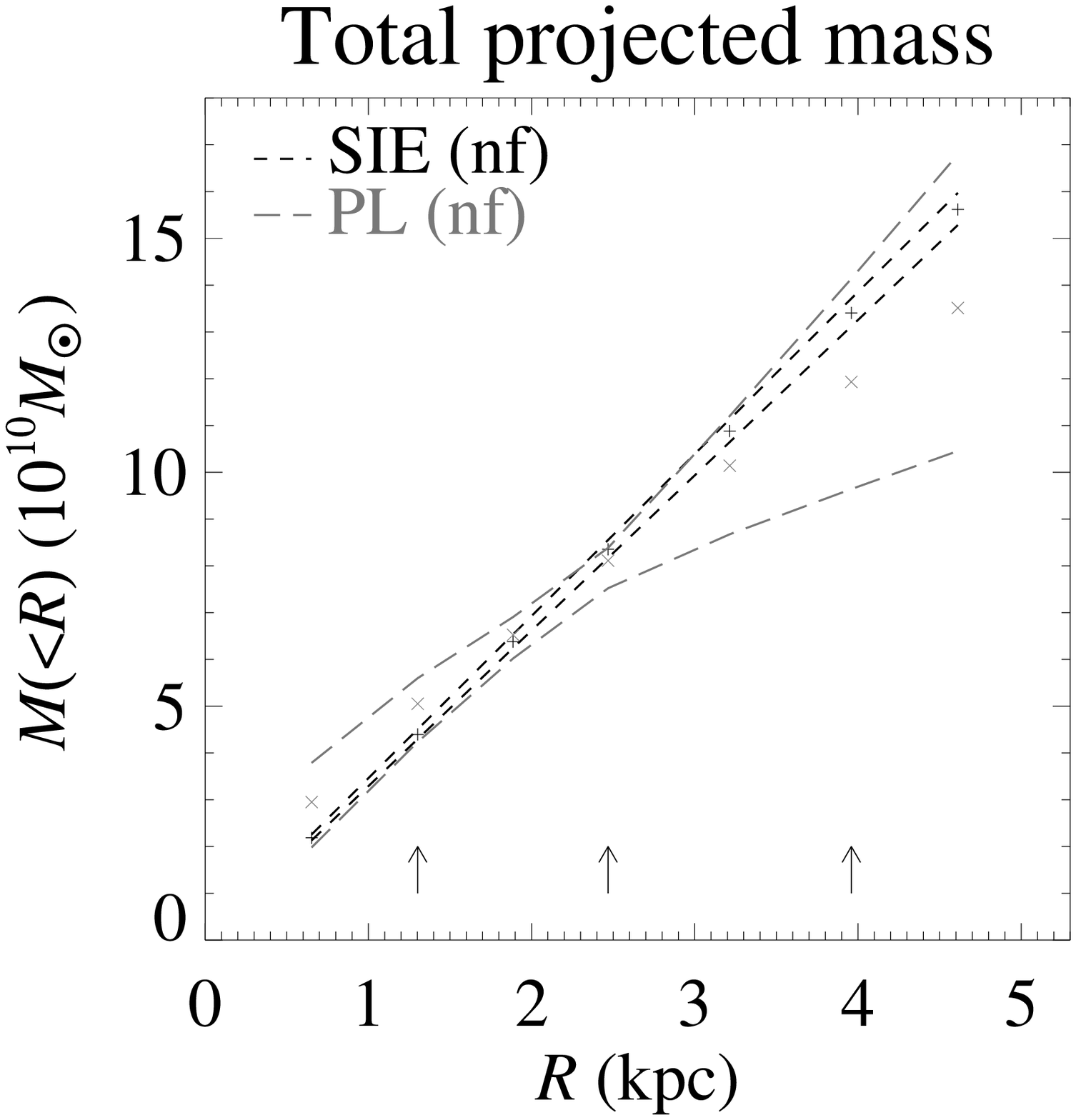}
\caption{Comparison of the projected total mass estimates for the SIE (nf), SIE (wf), and PL (nf) models. For each aperture, the 1$\sigma$ confidence intervals are determined from 2000 Monte-Carlo simulations by excluding the 320 smallest and the 320 largest mass estimates.} The arrows show the projected distances of the observed multiple images from the lens center.
\label{fi07}
\end{figure}


In Figs. \ref{fi05} and \ref{fi06}, we plot, respectively, the joint probability distributions of the SIE (nf) and PL (nf) model parameters, with the 68\% and 95\% confidence regions and the 68\% confidence intervals. These intervals are determined by excluding from the 2000 $\chi^{2}$ minimizations the 320 smallest and the 320 largest values for each model parameter. We have checked that the error estimates determined in this way are unbiased and equivalent to the uncertainties provided by a full Markov chain Monte Carlo analysis. The comparison Figs. \ref{fi05} and \ref{fi06} shows clearly that adding the exponent of the total mass distribution among the parameters increases their degeneracies, hence their error estimates. The probability distribution of the position angle $\theta_{q}$ is bimodal, with a secondary peak located nearly 90$^{\circ}$ away from the primary one, found at approximately 150$^{\circ}$. From the last column of plots in Fig. \ref{fi05}, we see that the secondary peak is included only in the 95\% CL regions, and, from the last panel of the same figure, we note that the low values of $\theta_{q}$ are associated with almost circular models (i.e., $q \simeq 1$). The bimodal distribution of the values of the lens position angle can then be explained by looking at the source plane of Fig. \ref{fi04}. If the axis ratio of a model is close to one, the shape of the tangential caustic is approximately symmetric with respect to the center of the lens and the surface enclosed by this caustic is small. In the same limit, the radial caustic is well approximated by a circle centered on the lens center. From these considerations, it follows that the expected positions of the images are almost invariant under a rotation of 90$^{\circ}$ of the lens mass distribution (supposing the positions of the sources are fixed). 

The degeneracies between $b$, $q$, and $\gamma$ are connected to their relations defined in Eqs. (\ref{eq01}) and (\ref{eq02}). In particular, the strong anti-correlation between the value of the length scale and the steepness (see the fifth panel of Fig. \ref{fi06}) is caused by the fact that the Einstein ring of a circular lens model defines a region on the image plane within which the average value of the convergence $\kappa$ is equal to one. In order for this equality to be approximately valid inside the average circle defined by the positions of our quad system, from Eq. (\ref{eq01}) and by holding the value of $q$ fixed, it follows that a higher value of $b$ requires a lower value of $\gamma$, and vice versa. 

The previous considerations on the almost model-independent average value of $\kappa$ inside the Einstein ring can also be translated in terms of total mass estimates within the same ring. Distinct models, defined by different parameters, that can reproduce well an approximately complete Einstein ring, provide total mass measurements inside this typical aperture that differ by only a few percent. This is shown in Fig. \ref{fi07}. There, we plot the median values and the 68\% confidence intervals (obtained by excluding from the 2000 $\chi^{2}$ minimizations the 320 smallest and the 320 largest mass estimates) of the lens projected total mass within the Einstein ring and measure values of $8.35_{-0.18}^{+0.20} \times 10^{10} M_{\odot}$ for an SIE (nf) model and $8.11_{-0.59}^{+0.27} \times 10^{10} M_{\odot}$ for a PL (nf) model. We notice that the median values of the 2000 Monte-Carlo cumulative total mass estimates do not necessarily follow a global PL model, but they have in principle more freedom. In fact, even if the total mass values of each of the 2000 models do follow a power law model precisely at all radii, the median values shown in Fig. \ref{fi07} and used in the following for the luminous and dark matter decomposition are more general and do not provide the same value of the steepness $\gamma$ at each radial position. In general, for the two different models the total mass estimates, that are measured within various apertures (approximately between 1 and 4 kpc from the lens center), are consistent, given the errors. We remark that fixing the exponent of the total mass profile (i.e., $\gamma$ equal to 2 for the SIE models) result in significant smaller uncertainties in the total mass values. As mentioned above, by modeling also the fluxes of the double system we find total mass measurements that are consistent within 1$\sigma$ with the estimates obtained by fitting the image positions only.

We generalize our result by emphasizing that the adoption of an isothermal model for strong lenses often provides a good fit of the observed images, but the errors on the projected mass estimates may be considerably underestimated already at projected distances from the center of the lens that differ from the Einstein radius by half its value. This fact has non-negligible consequences on the inferred properties of a lens dark matter distribution (see below). 

We notice that the value of $205 \pm 13$ km s$^{-1}$ for the central stellar velocity dispersion $\sigma_{0}$, which is determined by rescaling the value of the SDSS spectroscopic stellar velocity dispersion measured inside an aperture of 1.5\arcsec$\,$ [$\sigma = (189 \pm 12)$ km s$^{-1}$] to an aperture of radius equal to $\theta_{e}/8$, is consistent, within the errors, with the value of $215 \pm 5$, which is obtained by converting the total mass estimates shown in Fig. \ref{fi07} for the SIE (nf) model into an effective velocity dispersion $\sigma_{\mathrm{SIE}}$.

We remark that the best-fit parameters of our SIE point-like models are consistent, given the errors, with the best-fit parameters of the SIE extended model measured by \citet{bol08}. We also note that previous studies (\citealt{koc93,koc94}; \citealt{tre06}; \citealt{gri08b}) agree on finding that the central stellar velocity dispersion of early-type galaxies is a good estimator of the velocity dispersion of a one-component isothermal model.






\section{Luminous and dark matter}

We combine the surface brightness distribution measurement obtained from the HST images (see Sect. 2) with the multicolor photometric observations of the SDSS (see Table 1) and the projected total mass estimates determined from the lens modeling (see Sect. 3) to study the amount and distribution of luminous and dark matter in the lens galaxy.

\begin{figure}[htb]
\centering
\includegraphics[width=0.225\textwidth]{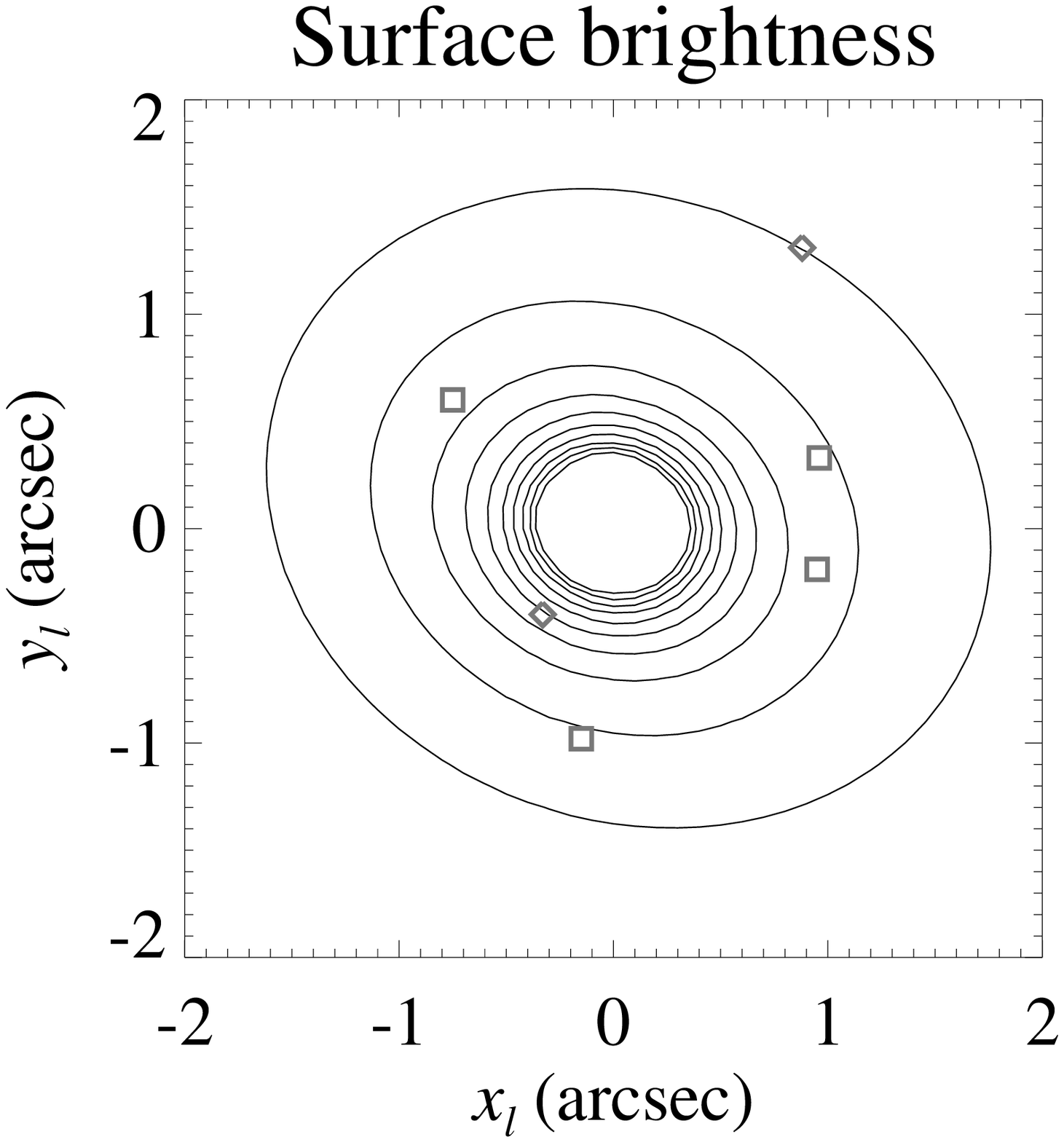}
\includegraphics[width=0.225\textwidth]{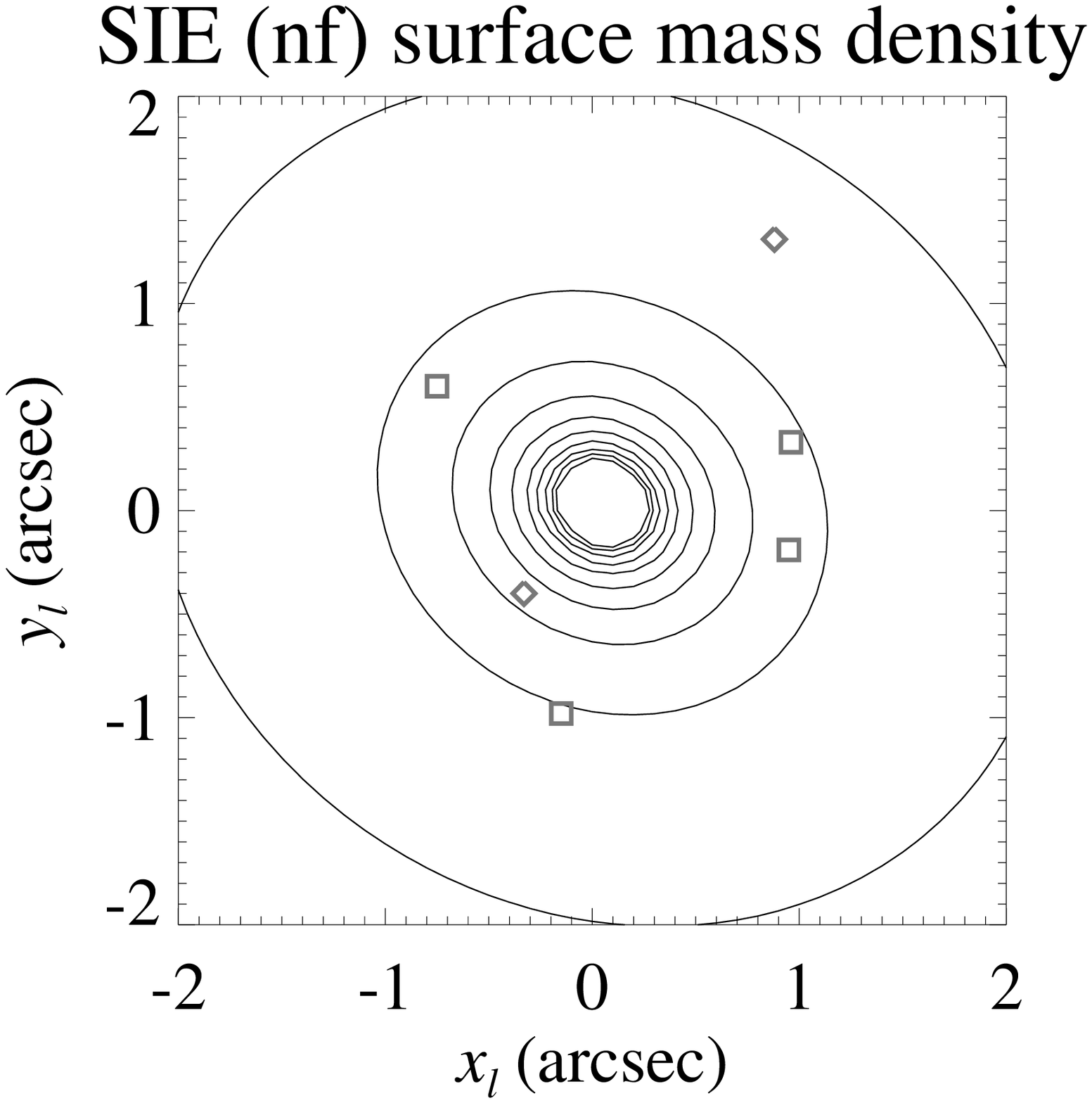}
\caption{Light and mass distributions. Isodensity contours of the best-fit surface brightness (\emph{on the left}) and parametric [SIE (nf), \emph{on the right}] total surface mass profiles. The observed image positions of the double (diamond) and quad (square) systems are shown.}
\label{fi10}
\end{figure}

First, we compare in Fig. \ref{fi10} the surface brightness and the total surface mass [for the SIE (nf) model] isodensity contours of the best-fit models described in the previous section. We use two images with the same area and pixel size, normalize the images to the sum of the values of all their pixels, and plot the same contour levels in both images. We observe that the distributions of light and total mass from the best-fit SIE (nf) model are nearly axisymmetric, but the former is slightly more concentrated than the latter. This can be inferred by looking at the positions of the inner and outer contour levels. The position angle of the surface brightness and total surface mass distributions are consistent within the errors. Thus, we conclude that the light distribution is approximately circular symmetric in projection and it is a good tracer of the total mass distribution.

\begin{figure}[!htb]
\centering
\includegraphics[width=0.49\textwidth]{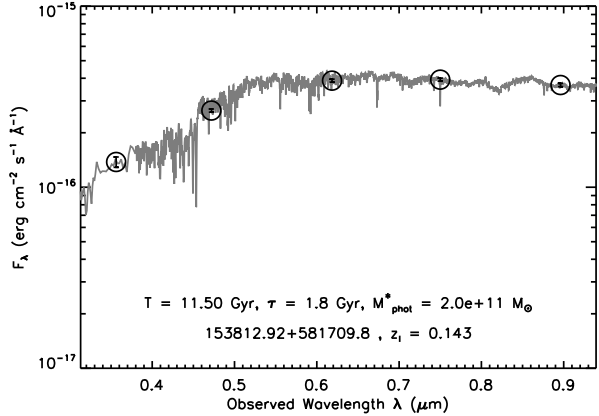}
\caption{SED and best-fit model of the lens galaxy SDSS J1538+5817. The observed total flux densities, measured in the \emph{u}, \emph{g}, \emph{r}, \emph{i}, and \emph{z} passbands, and their 1$\sigma$ errors are represented by circles and error bars. The best-fit is obtained by using Bruzual \& Charlot 2003 models. On the bottom, the best-fit values of the age ($T$), the characteristic time of the SFH ($\tau$), and the luminous mass ($M^{*}_{\mathrm{phot}}$) are shown.}
\label{fi11}
\end{figure}

\begin{figure}[!htb]
\centering
\includegraphics[width=0.22\textwidth]{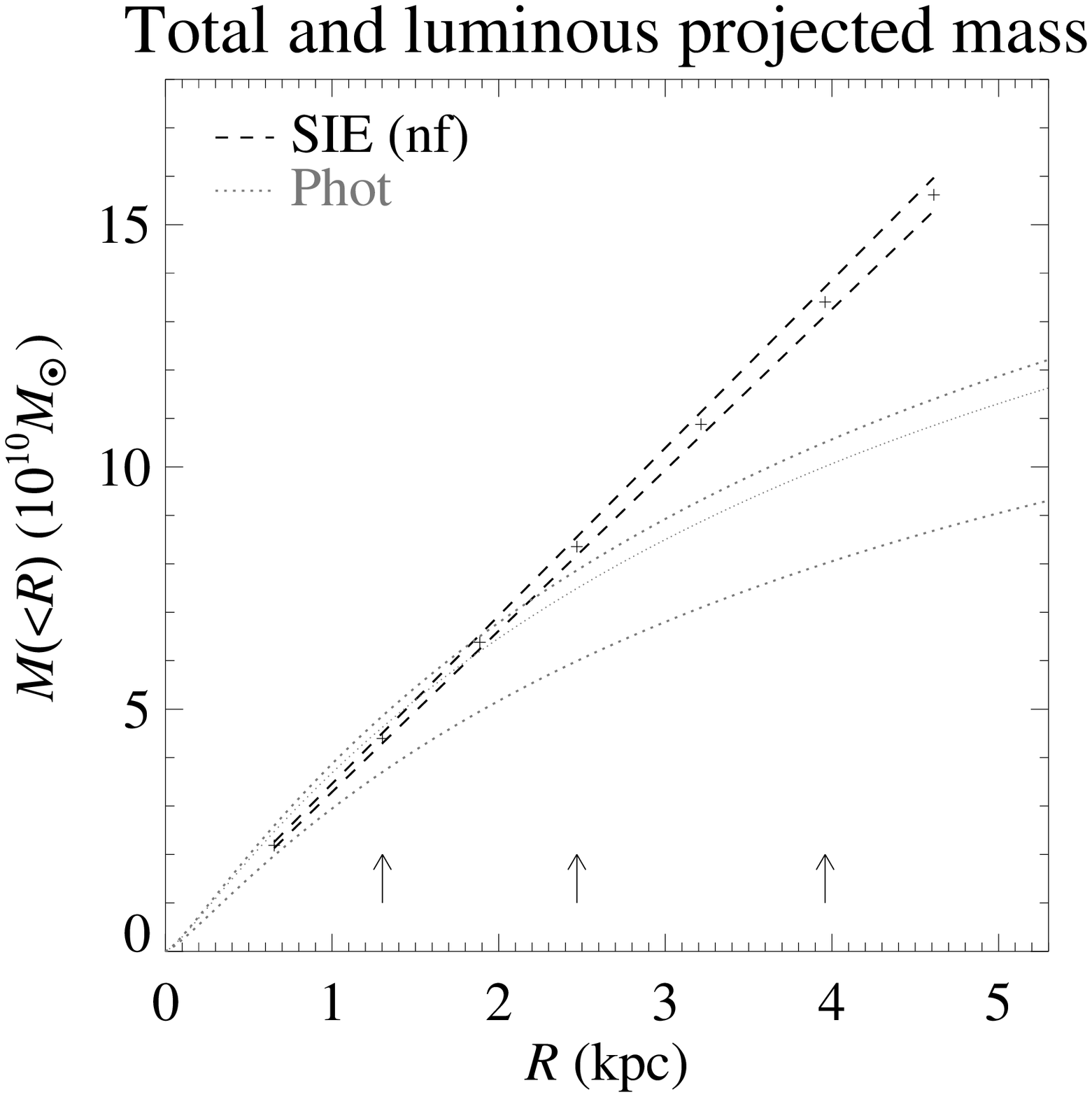}
\includegraphics[width=0.22\textwidth]{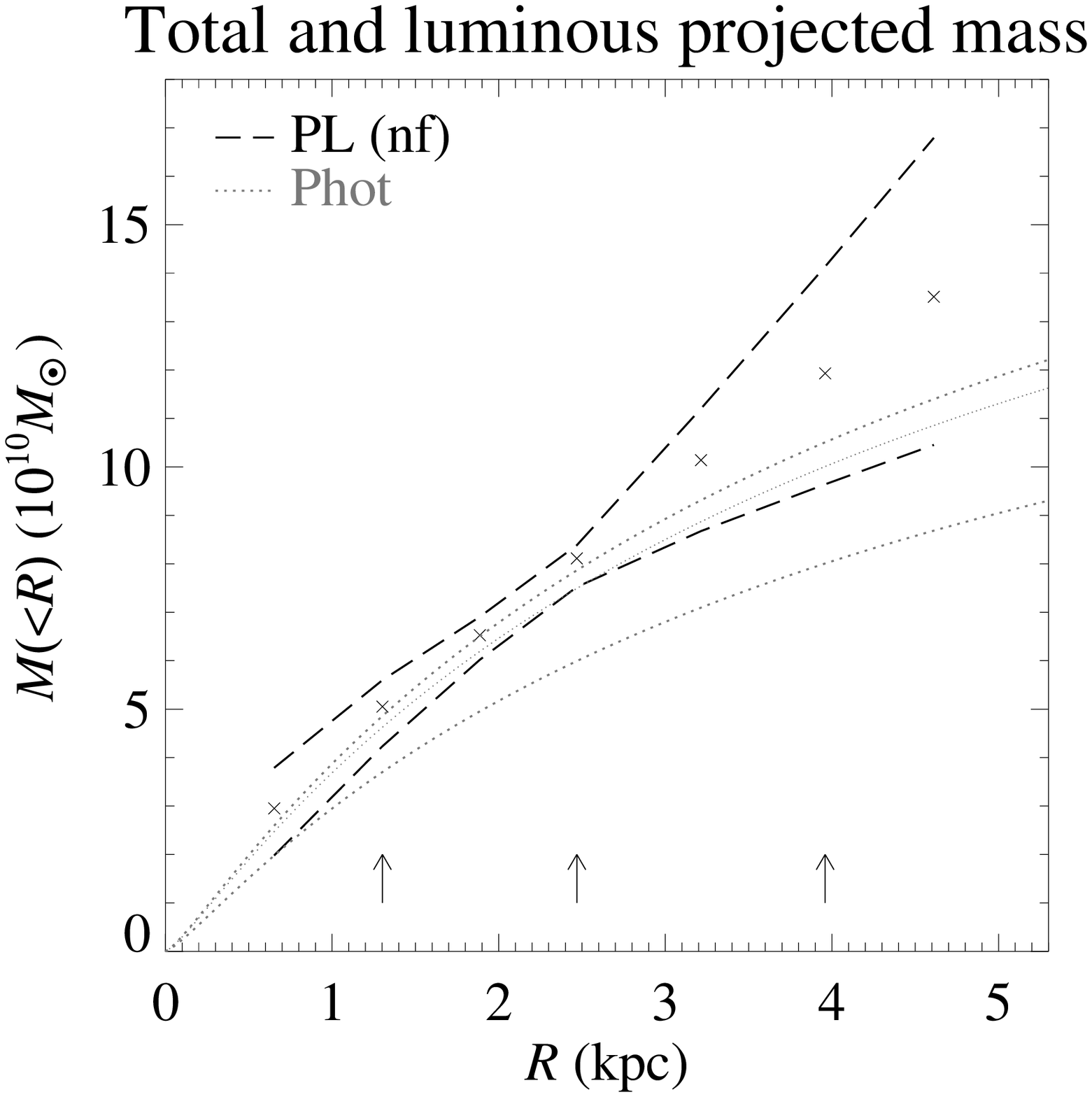}
\includegraphics[width=0.22\textwidth]{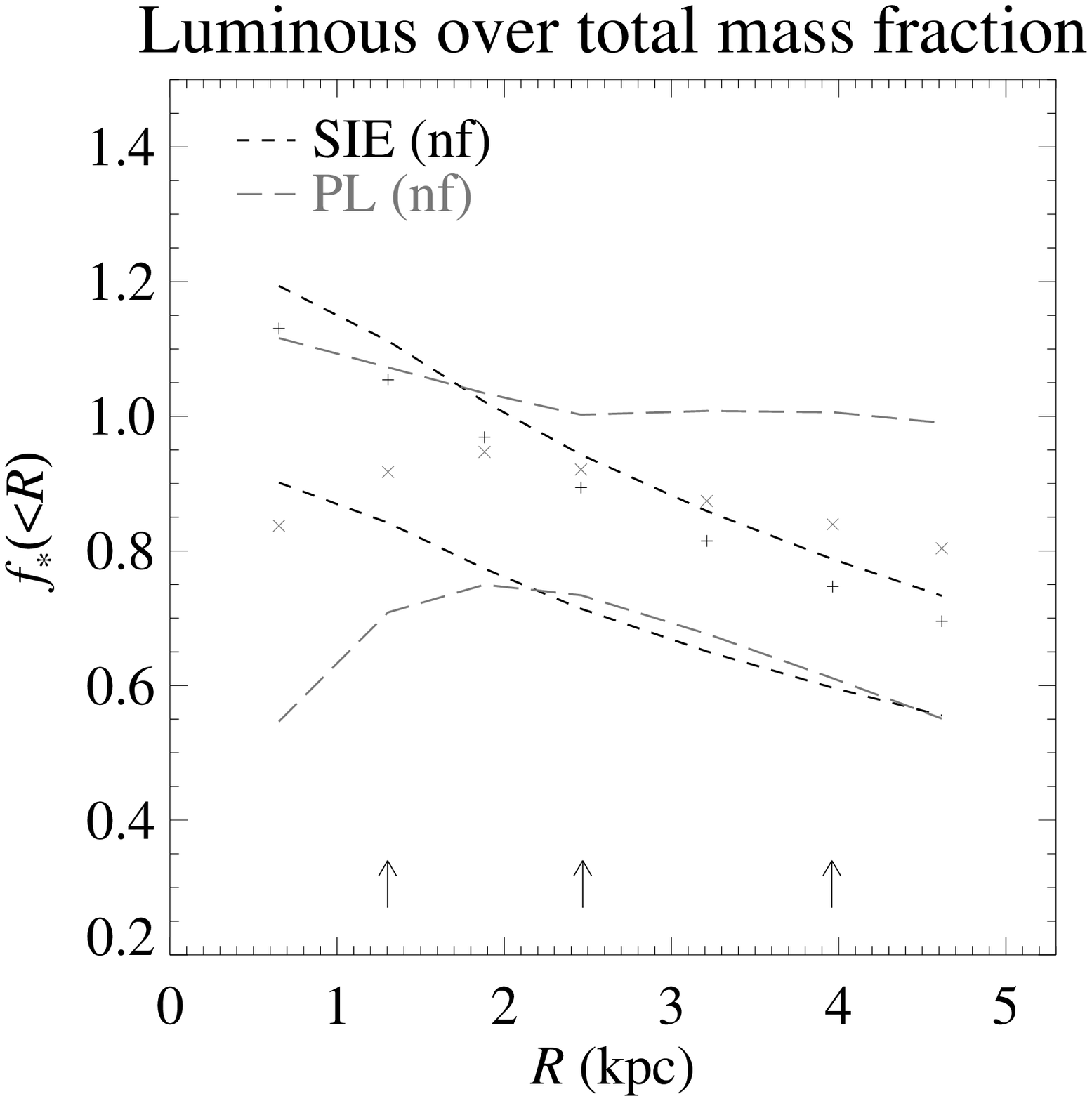}
\includegraphics[width=0.22\textwidth]{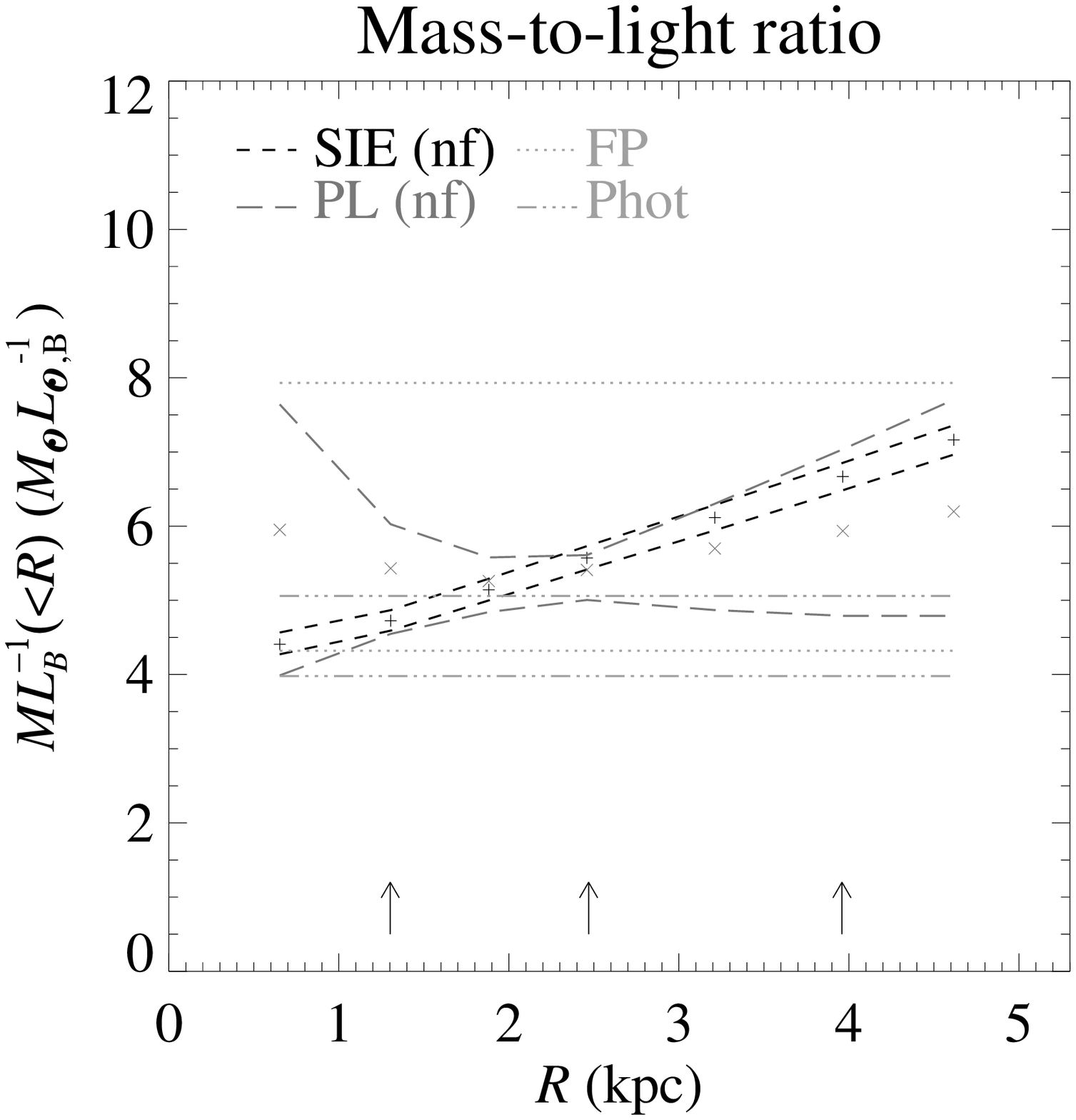}
\caption{Comparison of the projected total mass estimates of the SIE (nf) (\emph{on the top left}) and PL (nf) (\emph{on the top right}) models with the luminous mass measurements from the best-fit SED model. The fraction of projected mass in the form of stars (\emph{on the bottom left}) and the total and stellar mass-to-light ratios (\emph{on the bottom right}) are obtained by using lensing and photometric information. In each panel, the arrows show the projected distances of the observed multiple images from the lens center and the curves the 1$\sigma$ confidence intervals.}
\label{fi12}
\end{figure}

Next, we fit the lens spectral energy distribution (SED), consisting of the SDSS \emph{ugriz} magnitudes (see Table 1), with a three-parameter Bruzual \& Charlot composite stellar population (CSP) model computed by adopting a Salpeter initial mass function (IMF) and solar metallicity (for further details, see Grillo et al. 2009). The best-fit model, shown in Fig. \ref{fi11}, provides a photometric (luminous) mass $M_{\textrm{phot}}^{*}$ of the lens of $20^{+1}_{-4} \times 10^{10}\,M_{\odot}$. We then estimate the value of the mass in the form of stars $M^{*}_{\textrm{phot}}(\leq R)$, at a projected distance $R$ from the center of the lens, by multiplying $M_{\textrm{phot}}^{*}$ by an aperture factor $f_{\textrm{ap}}(\leq R)$, that represents the fraction of light measured within a circular aperture of radius $R$ divided by the total light of the galaxy. The quantities introduced above are explicitly defined as
\begin{equation}
M_{\textrm{phot}}^{*}(\leq  R) = M_{\textrm{phot}}^{*}\, f_{\textrm{ap}}(\leq  R)
\end{equation}
and
\begin{equation}
f_{\textrm{ap}}(\leq R) = \frac{\int_{0}^{R}I(\tilde{R})\tilde{R}\,\mathrm{d}\tilde{R}}{\int_{0}^{\infty}I(\tilde{R})\tilde{R}\,\mathrm{d}\tilde{R}},
\end{equation}
where $I(R)$ is the de Vaucouleurs profile
 \begin{equation}
I(R) = I_{0} \exp \Bigg[-7.67\bigg(\frac{R}{R_{e}}\bigg)^{\frac{1}{4}}\Bigg]\, ,
\end{equation}
with $R_{e}$ equal to $D_{\mathrm{ol}}\,\theta_{e}$. As discussed above, the circular symmetry of the light distribution assumed in the previous three equations is a plausible approximation for the lens surface brightness. In Fig. \ref{fi12}, we plot at different radii the projected total and luminous mass estimates obtained from the best-fit SIE (nf) and PL (nf) models of Sect. 3 and the best-fit SED model.

In the same figure, we show the fraction of projected mass in the form of stars 
\begin{equation}
f_{*}(\leq  R) := \frac{M_{\mathrm{phot}}^{*}(\leq  R)}{M_{\mathrm{lens}}^{\mathrm{tot}}(\leq  R)}
\end{equation}
and the total mass-to-light ratio 
\begin{equation}
\frac{M^{\textrm{tot}}}{L_{B}}(\leq  R) := \frac{M^{\textrm{tot}}_{\textrm{lens}}(\leq  R)}{L_{B}(\leq  R)}
\end{equation}
plotted versus the projected radius $R$. We compare this last quantity with the values of the stellar mass-to-light ratio estimated from the best-fit SED model ($M^{*}_{\textrm{phot}}\,L_{B}^{-1} = 5.0^{+0.1}_{-1.0}\,M_{\odot} L_{\odot,B}^{-1}$) and the evolution of the Fundamental Plane [$M^{*}_{\textrm{FP}} L_{B}^{-1} = (6.1 \pm 1.8)\,M_{\odot} L_{\odot,B}^{-1}$] (for more information, see Grillo et al. 2009).

The need for a dark component to be added to the luminous one to reproduce the total mass measurements of the SIE (nf) models is suggested by looking at the outer galaxy regions probed by lensing. Due to the larger error bars, the evidence on the presence of dark matter is reduced if the total mass estimates obtained from the PL (nf) models are considered. According to all the lensing models (i.e., deV, SIE, and PL), a value of $0.9^{+0.1}_{-0.2}$ for the fraction of projected mass in the form of stars over total is estimated at a projected distance from the galaxy center of approximately 2.5 kpc, and at 4 kpc from the galaxy center a value of one for the same quantity is excluded by the SIE (nf) models at more than 3$\sigma$ level. Moreover, at the same distance, the value of the total mass-to-light ratio determined from the SIE (nf) mass measurements is not consistent with the value of the mass-to-light ratio of the luminous component estimated from the galaxy SED modeling. Between 1 and 4 kpc, the same decrease of $f_{*}(\leq R)$ and deviation of $M^{\textrm{tot}}\,L_{B}^{-1}$ from $M_{\textrm{phot}}^{*}\,L_{B}^{-1}$ are also indicated by the values of the PL (nf) mass estimates, but these results are not highly significant because of the large uncertainties.

Finally, by taking advantage of the total mass measurements available at different distances from the center of the lens (not only in the vicinity of the Einstein angle, as in the majority of the known lensing systems), we decide to investigate the dark matter component in greater detail. To make possible a direct comparison of our results with those obtained from dynamical analyses or cosmological simulations, we consider two-component models in which the luminous $\rho_{\mathrm{L}}(r)$ and dark $\rho_{\mathrm{D}}(r)$ matter density distributions are parametrized by
\begin{eqnarray}
\rho_{\mathrm{L}}(r)&=&\frac{(3-\gamma_{\mathrm{L}})M_{\mathrm{L}}r_{\mathrm{L}}}{4\pi r^{\gamma_{\mathrm{L}}}(r+r_{\mathrm{L}})^{4-\gamma_{\mathrm{L}}}} \nonumber \\
\rho_{\mathrm{D}}(r)&=&\frac{(3-\gamma_{\mathrm{D}})M_{\mathrm{D}}r_{\mathrm{D}}}{4\pi r^{\gamma_{\mathrm{D}}}(r+r_{\mathrm{D}})^{4-\gamma_{\mathrm{D}}}}\,,
\label{eq:02}
\end{eqnarray}
where $M_{\mathrm{L/D}}$ is the total mass, $r_{\mathrm{L/D}}$ a break radius, and $\gamma_{\mathrm{L/D}}$ the inner density slope of the luminous and dark matter distributions.
The density profiles of Eq. (\ref{eq:02}) are projected along the line-of-sight to give the corresponding surface mass density profiles $\Sigma_{\mathrm{L/D}}(R)$:
\begin{eqnarray}
\Sigma_{\mathrm{L}}(R)&=&2\int_{R}^{\infty} \frac{\rho_{\mathrm{L}}(r)r\,\mathrm{d}r}{\sqrt{r^2-R^2}} \nonumber \\
\Sigma_{\mathrm{D}}(R)&=&2\int_{R}^{\infty} \frac{\rho_{\mathrm{D}}(r)r\,\mathrm{d}r}{\sqrt{r^2-R^2}}\,,
\label{eq:03}
\end{eqnarray}
which, once integrated, result in the following cumulative mass distributions $M_{\mathrm{L/D}}(\leq  R)$:
\begin{eqnarray}
M_{\mathrm{L}}(\leq  R)&=&\int_{0}^{R}\Sigma_{\mathrm{L}}(R)\,2\pi R \,\mathrm{d}R \nonumber \\
M_{\mathrm{D}}(\leq  R)&=&\int_{0}^{R}\Sigma_{\mathrm{D}}(R)\,2\pi R \,\mathrm{d}R \,.
\label{eq:04}
\end{eqnarray}
The total density $\rho_{\mathrm{T}}(r)$, surface mass density $\Sigma_{\mathrm{T}}(R)$, and cumulative mass $M_{\mathrm{T}}(R)$ distributions are defined as the sum of the luminous and dark contributions
\begin{eqnarray}
\rho_{\mathrm{T}}(r) &=& \rho_{\mathrm{L}}(r)+\rho_{\mathrm{D}}(r)\,, \nonumber \\
\Sigma_{\mathrm{T}}(R) &=& \Sigma_{\mathrm{L}}(R)+\Sigma_{\mathrm{D}}(R)\,, \nonumber \\
M_{\mathrm{T}}(\leq  R) &=& M_{\mathrm{L}}(\leq  R)+M_{\mathrm{D}}(\leq  R)\,.
\label{eq:05}
\end{eqnarray}
We notice again that the circular approximation is plausible for this particular lens. 

The luminous quantities introduced in the above equations are completely determined from the photometric observations. In fact, for the luminous component we have estimated the total mass $M_{\mathrm{L}}$ by modeling the SED and, to obtain a surface brightness profile close to a de Vaucoleurs profile, we assume a Hernquist (1990; $\gamma_{\mathrm{L}} = 1$ and $r_{\mathrm{L}}=R_{e}/1.8153$) or a Jaffe (1983; $\gamma_{\mathrm{L}} = 2$ and $r_{\mathrm{L}}=R_{e}/0.7447$) density profile. Then, we construct a grid of 13671 models for the dark component. The total mass $M_{\mathrm{D}}$, the break radius $r_{\mathrm{D}}$, and the inner density slope $\gamma_{\mathrm{D}}$ can assume values included between 0.1 and 100 times $M_{\mathrm{L}}$, 0.1 and 10\arcsec, and 0.5 and 2.5, respectively. The first two intervals are divided logarithmically into 31 and 21 points respectively, the last one linearly into 21 points. The best-fit dark matter profile is found by minimizing the following chi-square function:
\begin{equation}
\begin{scriptsize}
\chi^{2}(M_{\mathrm{D}},r_{\mathrm{D}},\gamma_{\mathrm{D}}) = \sum_{i=2}^{6} \Bigg[ \frac{ M_{\mathrm{lens}}^{\mathrm{tot}}(\leq  R_{i}) - M_{\mathrm{T}}(\leq  R_{i})} {\sigma_{M_{\mathrm{lens}}^{\mathrm{tot}}(\leq  R_{i}) }} \Bigg]^{2} \,.   
\end{scriptsize}
\end{equation}
In order to estimate the errors in the best-fit parameters, we perform 500 Monte-Carlo simulations varying the total mass of the luminous component according to the corresponding measurement errors and the luminous break radius by assuming a realistic 10\% uncertainty. 

\begin{figure}[!htb]
\centering
\includegraphics[width=0.45\textwidth]{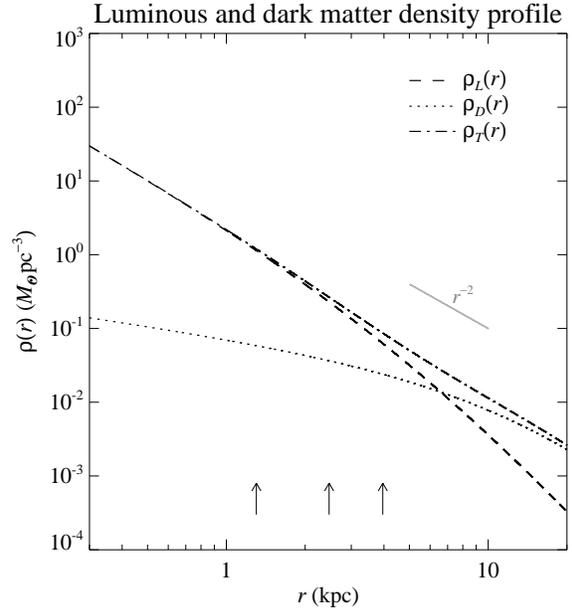}
\caption{Best-fit (minimum chi-square) luminous and dark matter decomposition, determined by assuming a Jaffe profile for the three-dimensional luminous density and projected total mass measurements as estimated from the PL (nf) model. The arrows show the projected distances of the observed multiple images from the lens center.}
\label{fi13}
\end{figure}

\begin{figure} [!htb]
\centering
\includegraphics[width=0.225\textwidth]{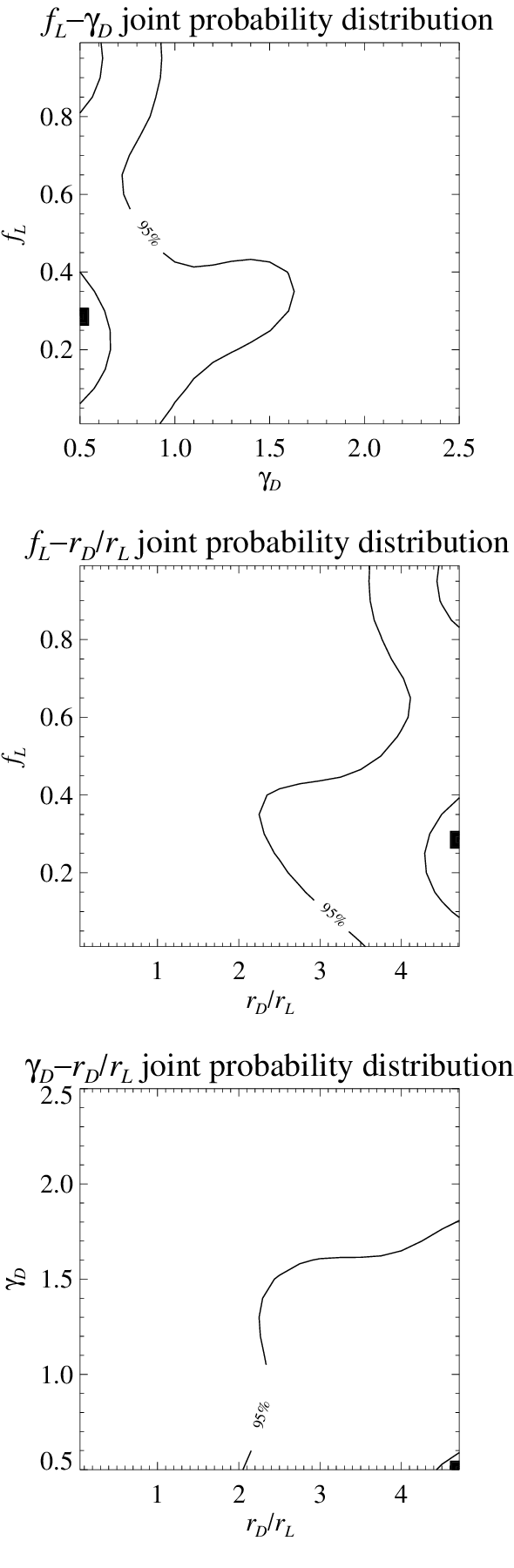}
\includegraphics[width=0.225\textwidth]{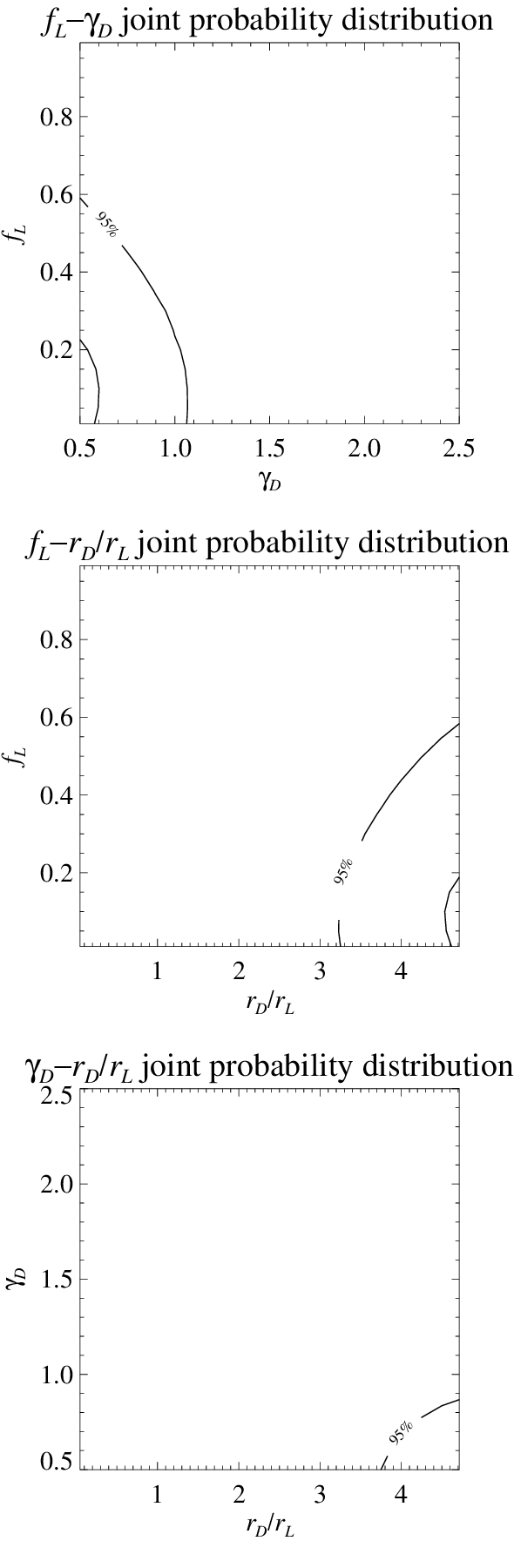}
\caption{Estimates of the errors and correlations in the parameters related to the dark matter component: the luminous mass fraction $f_{\mathrm{L}}$, the dark to luminous break radius ratio $r_{\mathrm{D}}/r_{\mathrm{L}}$, and the dark matter inner density slope $\gamma_{\mathrm{D}}$. The projected total mass measurements of the PL (nf) (\emph{on the left}) and SIE (nf) (\emph{on the right}) models are used. The small squares on the three left panels show the best-fit parameters corresponding to the dark matter density profile represented in Fig. \ref{fi13}.}
\label{fi14}
\end{figure}

In Fig. \ref{fi13}, we show the luminous and dark mass decomposition obtained from the best-fit (minimum chi-square) model and in Fig. \ref{fi14} the parameter joint probability distributions. We decide to plot the best-fit dark matter model obtained by assuming a Jaffe profile (no significative differences are present if a Hernquist profile is adopted) for the luminous component and considering the projected total mass estimates coming from the PL (nf) models. The confidence levels on the parameter space of the dark matter component are expressed in terms of the luminous mass fraction $f_{\mathrm{L}}=M_{\mathrm{L}}/(M_{\mathrm{L}}+M_{\mathrm{D}})$, i.e., the mass in the form of stars to the total mass of the galaxy, the ratio of the dark to luminous break radius $r_{\mathrm{D}}/r_{\mathrm{L}}$, and $\gamma_{\mathrm{D}}$. 

We find a best-fit $\chi^{2}$ value of 0.8 with two degrees of freedom (derived from the total mass measurements at the five central radii fitted by three-parametric models). We measure that the values of the dark matter density overcome those of the luminous matter density at radii larger than approximately 1.5 times the effective radius of the galaxy ($R_{e}$ = 4.0 kpc). As in the previous sections, a three-dimensional total density profile close but not exactly equal to a function decreasing as $1/r^{2}$ (i.e., an isothermal profile) is found. We note that the uncertainties in the dark matter parameters determined by using the projected total mass estimates of the PL (nf) models are significantly larger than those coming from the measurements of the SIE (nf) models. This is a consequence of the different error sizes of the two sets of projected total mass estimates. For the same reason, as already discussed looking at Fig. \ref{fi12}, large values of $f_{\mathrm{L}}$ are excluded at a 95\% CL only if the lens three dimensional total density profile is fixed to be isothermal. We observe that the dark matter component is in any case more diffused than the luminous one. In fact, $r_{\mathrm{D}}/r_{\mathrm{L}}$ is larger than 2 at more than a 95\% CL. Given the assumed parametrization, we also find that the dark matter density profile $\rho_{\mathrm{D}}(r)$ is probably shallow in the inner galactic regions. The value of $\gamma_{\mathrm{D}}$ is indeed lower than 0.7 at a 68\% CL.

\section{Summary and conclusions}

By means of HST/ACS and WFPC2 imaging and NOT/ALFOSC spectroscopy, we have established that SDSS J1538+5817 is a rare lensing system composed of a luminous elliptical galaxy, located at redshift $z_{l}=0.143$, that acts as a lens on two distinct and equally-distant ($z_{s}=0.531$) sources. The two sources are lensed into a double and a quad (with an almost complete Einstein ring) system, covering rather large angular and radial scales on the lens plane. This exceptional configuration has allowed us to investigate in great detail the lens total mass distribution within the effective radius of the galaxy, through parametric and non-parametric point-like lensing programs and perform a complete statistical study of the errors and correlations on the lens model parameters. Then, by fitting the lens SED with CSP models, we have estimated the luminous mass of the galaxy and combined the lensing and photometric measurements to examine the characteristics of the galaxy dark-matter halo.

In detail, our main results can be summarized in the following points:
\begin{itemize}

\item[$-$] Parametric models predict image positions that match closely the observed lensing geometry, and describe lens total mass distributions that are almost circular in projection, moderately steeper than an isothermal profile, and well aligned with the lens light distribution.

\item[$-$] The value of the total mass projected within the Einstein circle of radius 2.5 kpc is slightly larger than $8 \times 10^{10}\,M_{\odot}$ and approximately 10\% of this mass is in the form of dark matter.

\item[$-$] In the inner galactic regions, the galaxy dark-matter density distribution is shallower and more diffuse than the luminous one. The former starts exceeding the latter at a distance of roughly 6 kpc from the galaxy center, corresponding to 1.5 times the value of the luminous effective radius.

\end{itemize}
  
We conclude by remarking that strong gra\-vi\-ta\-tio\-nal lens systems with configurations comparable to or more complex than that of SDSS J1538+5817 are excellent laboratories to study the distribution of luminous and dark matter in early-type galaxies. However, to achieve realistic results on the dark matter component, it is essential to verify the commonly accepted isothermality of the total mass distribution at a higher level than done so far. Strong lensing systems with an Einstein radius significantly larger than the effective radius of the lens galaxy would be invaluable to determine the dark matter properties of the halos of early-type galaxies.





\acknowledgments

We acknowledge the support of the European DUEL Research Training Network, Transregional Collaborative Research Centre TRR 33, and Cluster of Excellence for Fundamental Physics and the use of data from the accurate SDSS database. We thank Piero Rosati for useful suggestions and the NOT staff for carrying out our observations in service mode. We are grateful to the NOT Scientific Association for awarding some observing time solely on the basis of scientific merit and supporting the NOT Summer School where CG gained some observational experience with the ALFOSC spectroscopic data.  






\appendix
\section{Non-parametric models}
\label{sec:non-parametric models}

\emph{PixeLens} (\citealt{sah04}) is a non-parametric lensing program that generates an ensemble of models consistent with the observed data of a lensing system. Each model is composed of a pixelated surface mass density map of the lens, the reconstructed position of the source, and, optionally, an estimate of the value of the Hubble parameter. These results are obtained by using the observed positions of the multiple images (ordered by arrival time, even if time delays are not known), the redshifts of the lens and the source, and some priors based on previous knowledge of general galaxy mass distribution (for further details, see \citealt{sah97}; \citealt{col08}). Interestingly, \emph{PixeLens} has been employed to measure the value of the Hubble parameter from samples of strong lensing systems with measured time delays (e.g., \citealt{sah06}; \citealt{col08}). 

We model here the surface mass density of the lens on a symmetric circular grid of 2\arcsec $\,$ radius divided into 20 pixels. We consider 400 models with fixed cosmological values and with decreasing total projected mass profiles [i.e., $\Sigma(R) \propto R^{-\alpha}$, where $\alpha > 0$].

\begin{figure*}[!htb]
\centering
\includegraphics[width=0.22\textwidth]{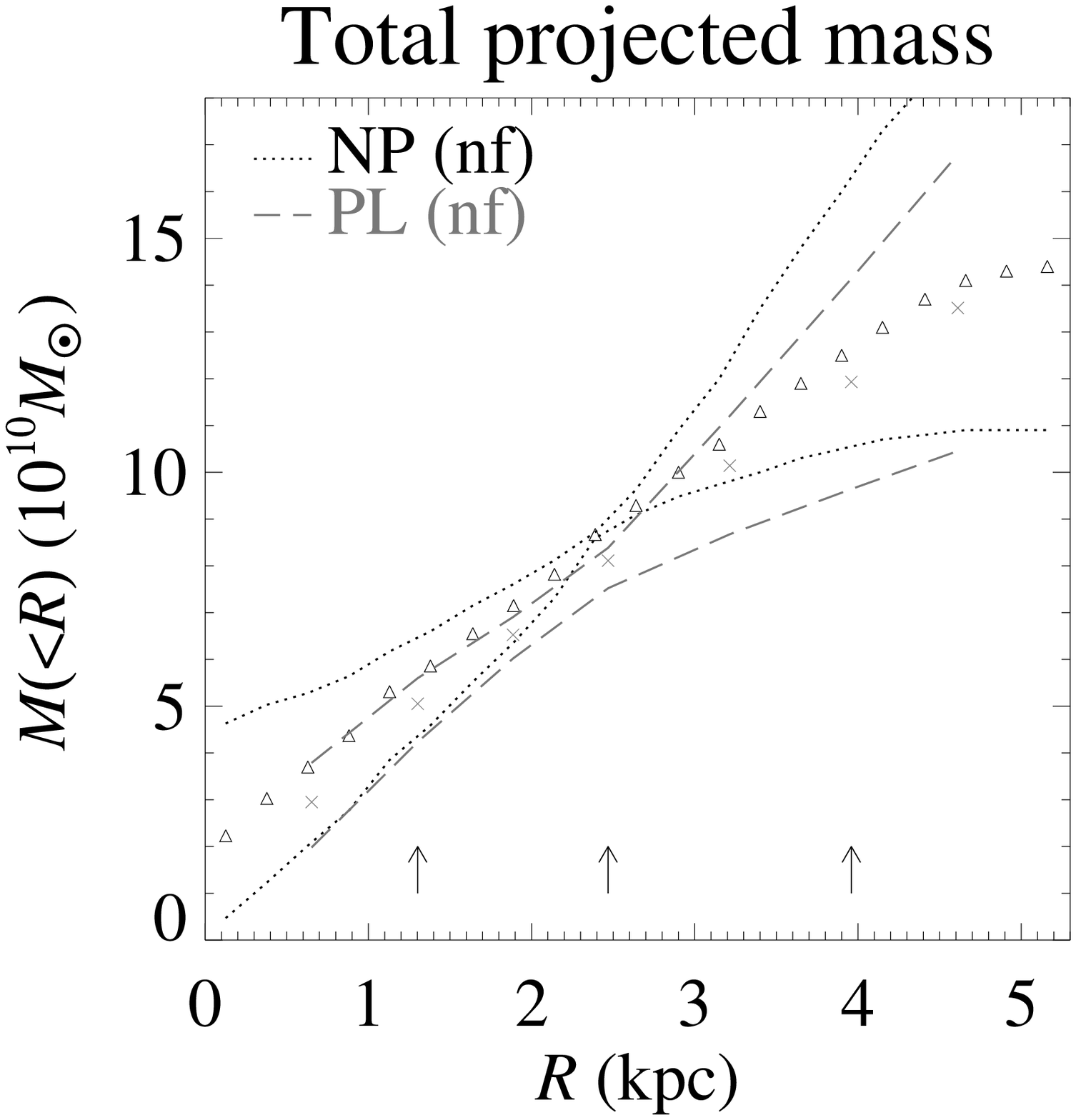}
\includegraphics[width=0.22\textwidth]{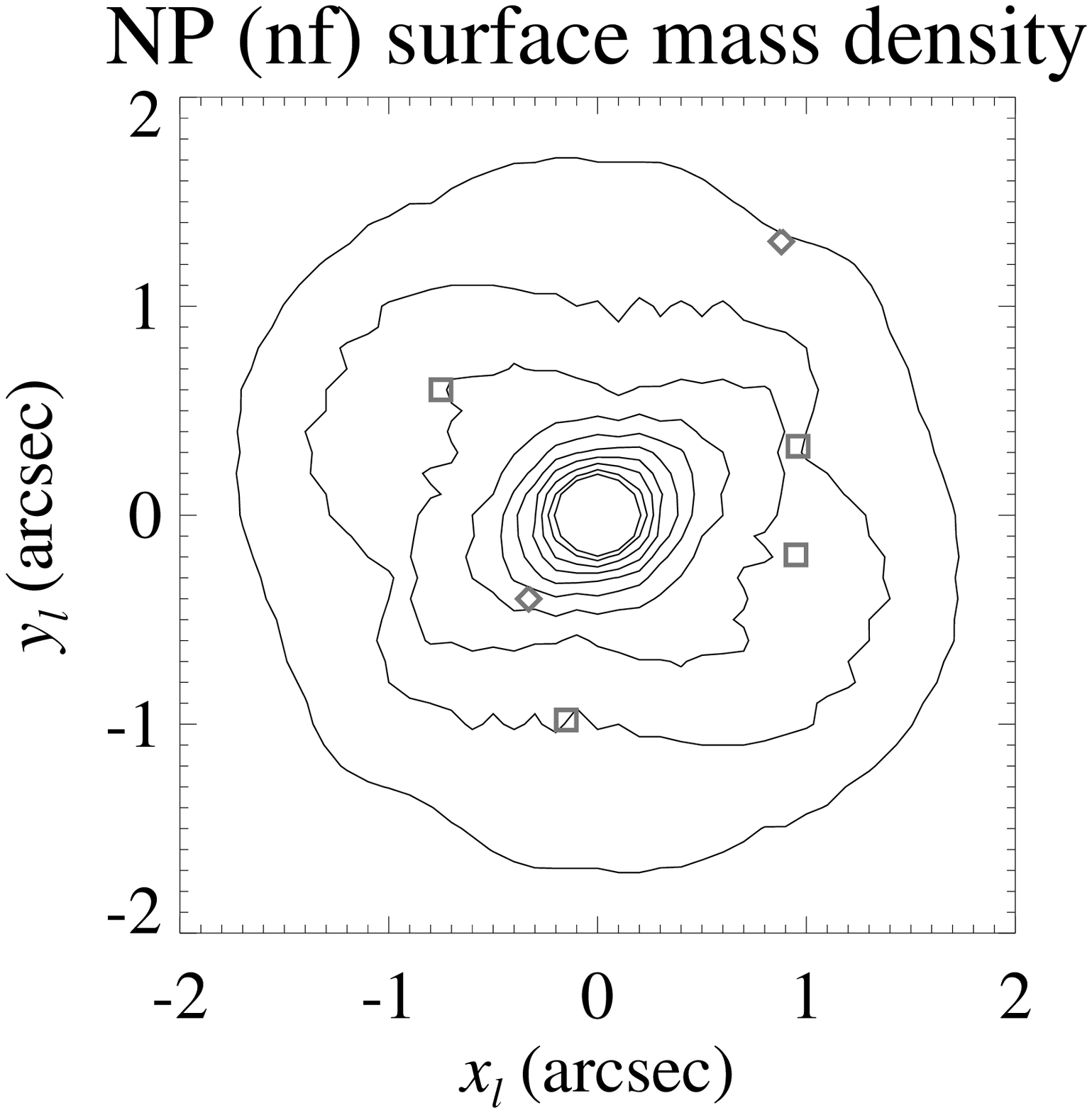}
\includegraphics[width=0.22\textwidth]{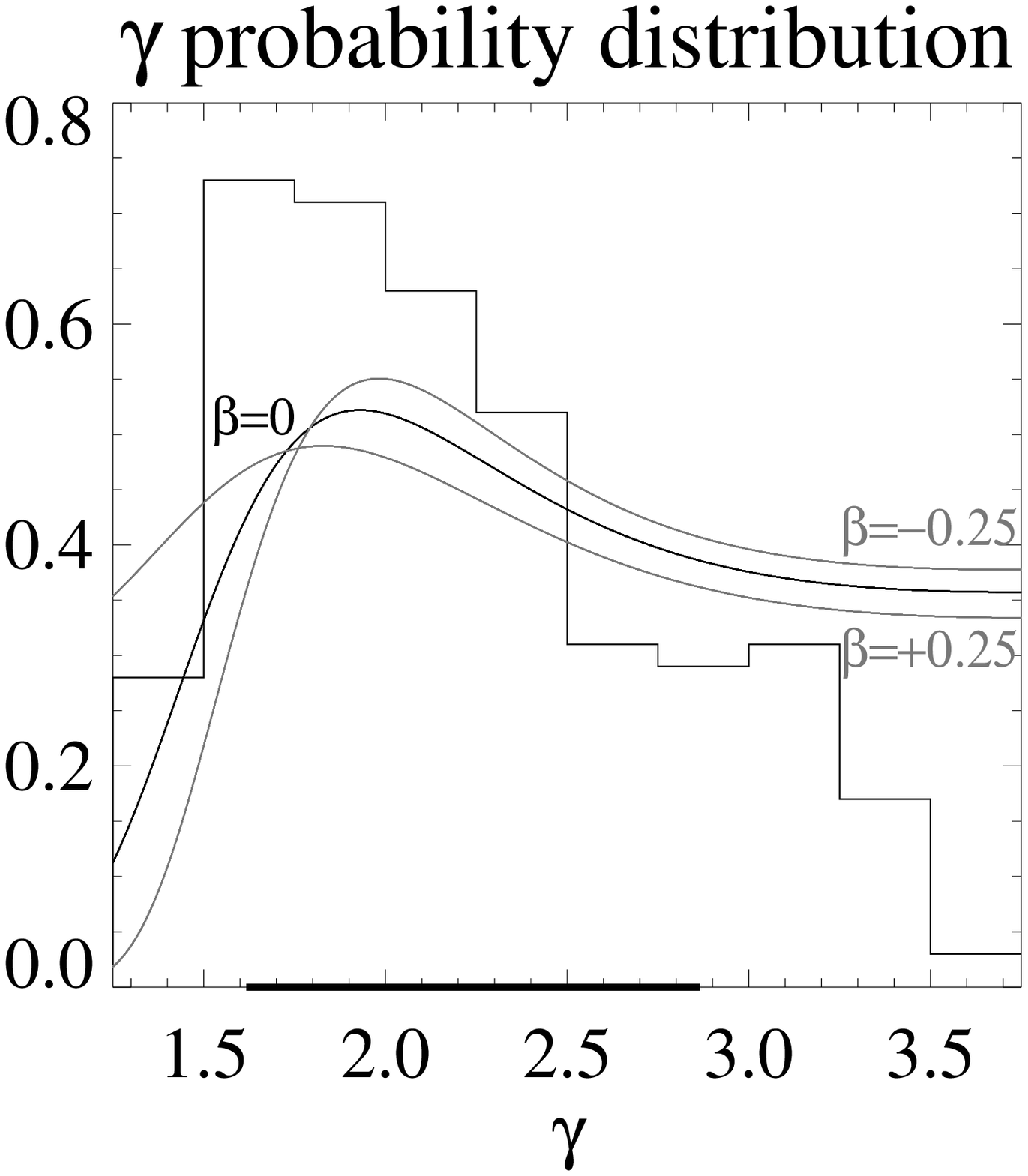}
\caption{\emph{Left:} Comparison of the total projected mass estimates, with 1$\sigma$ confidence intervals, from parametric [PL (nf)] and non-parametric [NP (nf)] modeling. The arrows show the projected distances of the observed multiple images from the lens center. \emph{Middle:} Isodensity contours of the best-fit non-parametric [NP (nf)] total surface mass profile. The observed image positions of the double (diamond) and quad (square) systems are shown. \emph{Right:} Marginal probability distribution (histogram) of the three-dimensional total density exponent $\gamma$ from non-parametric modeling. The thick bar on the \emph{x}-axis shows the 1$\sigma$ confidence interval. The same probability distribution as obtained by combining strong lensing and stellar dynamics measurements is represented by the smooth curves.}
\label{fi09}
\end{figure*}

The cumulative total projected mass, the total surface mass density profile of the average model, and the marginalized probability distribution of the three-dimensional total density exponent $\gamma$ are shown in Fig. \ref{fi09}. We measure a value of the total mass projected within the Einstein radius of $8.59^{+0.13}_{-0.12} \times 10^{10} M_{\odot}$, at a 68\% CL. At the same confidence level, we estimate a value of $\gamma$ included between 1.62 and 2.87. We observe that the contour levels of the non-parametric total surface mass show non-negligible values of ellipticity in the inner regions. The differences between the surface brightness of Fig.~\ref{fi10} and total surface mass of  Fig.~\ref{fi09} are significant within the area defined by the Einstein radius. This is not surprising since here the total surface mass density distribution is almost completely unconstrained by the lensing observables. This is the equivalent of Gauss' law in gravitational lensing (see \citealt{koc04}). We notice that these differences are less evident outside the Einstein ring, where the positions of the multiple images limit the freedom of the non-parametric models in determining the lens total mass distribution. In Fig. \ref{fi09}, we also show for comparison the mass estimates obtained in the equivalent parametric modeling [PL (nf)] and the probability distribution of the density exponent that is expected by combining strong lensing and stellar dynamics measurements. In detail, the combined lensing and dynamical probability distribution for $\gamma$ is obtained by using the following expression
\begin{equation}
\frac{c^{2}}{4\pi}\frac{\theta_{\mathrm{Ein}}}{\sigma_{0}^{2}}\,\tilde{r}(z_{l},z_{s};\Omega_{m},\Omega_{\Lambda})=\bigg(\frac{8\theta_{\mathrm{Ein}}}{\theta_{e}}\bigg)^{2-\gamma}g(\gamma,\delta,\beta)   
\end{equation}
that relates through the spherical Jeans equations the values of the central stellar velocity dispersion $\sigma_{0}$, Einstein angle $\theta_{\mathrm{Ein}}$, effective angle $\theta_{e}$, exponent of the three-dimensional luminosity density profile $\delta$, anisotropy parameter of the stellar velocity ellipsoid $\beta$, and ratio of angular diameter distances between observer-source and lens-source $\tilde{r}(z_{l},z_{s};\Omega_{m},\Omega_{\Lambda})$ [$g(\gamma,\delta,\beta)$ is a numerical factor that depends on the three cited quantities; for definitions and further details, see \citealt{koo05a}]. In the plots of Fig.\ref{fi09}, we fix $\delta$ equal to 2 and choose two values of $\beta$ ($-0.25$ and $+0.25$) representative of small tangential and radial orbit anisotropy. By doubling the size of the grid but keeping the same size of the pixels, we have checked that the choice of a circular grid with a radius of 2\arcsec $\,$ to reconstruct the total surface mass density distribution of our not perfectly circular lens galaxy does not introduce any artificial shear component and does not affect significantly the results.

According to these results and looking at Fig. \ref{fi09}, we can conclude that the two independent parametric and non-parametric analyses are in general consistent, within the errors, as far as total projected mass and three-dimensional total density exponent measurements are concerned, but small differences and some considerations are worth noticing. 

The projected total mass estimates obtained with \emph{PixeLens} are systematically larger than those obtained with \emph{gravlens}. This can be caused by a combination of the mass-sheet degeneracy (see \citealt{fal85}; \citealt{sch95}) and the prior on the positive definiteness of every pixel of the grid of the total surface mass density. Among all the arbitrary constants that can be added to the convergence $\kappa$, leaving though the image positions unchanged, those which provide a negative value of $\kappa$ somewhere on the grid are excluded, \emph{a priori}, from the non-parametric lensing analysis. This fact may bias the projected total mass measurements to slightly larger values.

As far as $\gamma$ is concerned, the larger uncertainty coming from the non-parametric reconstruction with respect to the parametric one is probably just a consequence of the more general allowed models. A bias towards small values of $\gamma$ may be associated to the prior present in \emph{PixeLens} that constrains the value of $\kappa$ on one pixel of the grid to be lower than twice the average value of the neighboring pixels. For large values of $\gamma$, two adjacent pixels located in the central region of the lens may have very different values of $\kappa$, hence these models may not be included in the statistical ensemble.

Finally, we remark on the overall agreement between the lensing only and lensing plus dynamics probability distributions of $\gamma$. We notice, though, that lensing alone does not reach the precision needed to distinguish among models with different values of the stellar anisotropy parameter $\beta$.

\clearpage




\clearpage

\clearpage

\end{document}